\documentclass[aps,prd,reprint,superscriptaddress,fleqn]{revtex4-1}
\usepackage{graphicx}
\usepackage{dcolumn}
\usepackage{natbib,aas_macros}
\usepackage{color}
\usepackage{subfig}
\usepackage{mathtools}
\usepackage{amsmath}
\usepackage{caption}
\usepackage{soul}

\newcommand{\add}[1]{\textcolor{black}{#1}}
\newcommand{\km}[1]{\textcolor{black}{#1}}

\newcommand{\bz}[1]{\textcolor{black}{#1}}

\begin{document}
\title{Energetics of ultrahigh-energy cosmic-ray nuclei}

\author{Yu Jiang}
\affiliation{Department of Physics, The Pennsylvania State University, University Park, Pennsylvania 16802, USA}
\affiliation{Department of Astronomy \& Astrophysics, The Pennsylvania State University, University Park, Pennsylvania 16802, USA}
\affiliation{Center for Multimessenger Astrophysics, Institute for Gravitation and the Cosmos, The Pennsylvania State University, University Park, Pennsylvania 16802, USA}
\author{B. Theodore Zhang}
\affiliation{Department of Physics, The Pennsylvania State University, University Park, Pennsylvania 16802, USA}
\affiliation{Department of Astronomy \& Astrophysics, The Pennsylvania State University, University Park, Pennsylvania 16802, USA}
\affiliation{Center for Multimessenger Astrophysics, Institute for Gravitation and the Cosmos, The Pennsylvania State University, University Park, Pennsylvania 16802, USA}
\author{Kohta Murase}
\affiliation{Department of Physics, The Pennsylvania State University, University Park, Pennsylvania 16802, USA}
\affiliation{Department of Astronomy \& Astrophysics, The Pennsylvania State University, University Park, Pennsylvania 16802, USA}
\affiliation{Center for Multimessenger Astrophysics, Institute for Gravitation and the Cosmos, The Pennsylvania State University, University Park, Pennsylvania 16802, USA}
\affiliation{Center for Gravitational Physics, Yukawa Institute for Theoretical Physics, Kyoto, Kyoto 606-8502 Japan}

\date{\today}

\begin{abstract}
Energetics of the ultrahigh-energy cosmic rays (UHECRs) generated in the universe is crucial for pinning down their candidate sources.
Using the recent Auger data on UHECR spectra, we calculate the UHECR energy generation rate density for different species of nuclei at the injection, considering intermediate and heavy nuclei as well as protons, through scanning over source parameters on the spectral index, maximum energy and redshift evolution. 
We find the resulting UHECR energy generation rate density to be $\approx(0.2-2)\times$10$^{44}$~erg~Mpc$^{-3}$~yr$^{-1}$ at $10^{19.5}$~eV with a nontrivial dependence on the spectral index. 
Nuclei other than protons and irons favor hard spectral indices at the injection, and the required value of energy budget is smaller for intermediate nuclei.
Given significant uncertainties in hadronic interaction models and astrophysical models for the Galactic-extragalactic transition, our results can be regarded as conservative.  
The composition data on $X_{\rm max}$ give additional constraints, but the results are consistent within the model uncertainties. 
\end{abstract}

\maketitle

\section{Introduction}
Cosmic rays with energies higher than $3\times10^{18}$eV are called ultrahigh-energy cosmic rays (UHECRs) that are the highest-energy particles known so far~\cite{Linsley:1963bk}.
The origin of UHECRs remains a half-century mystery despite tremendous efforts~\cite{Kotera:2011cp,Anchordoqui:2018qom,AlvesBatista:2019tlv}.
In particular, the Pierre Auger Observatory and Telescope Array (TA) are two largest cosmic-ray experiments up to date that have collected data towards investigating the nature and origin of UHECRs~\cite{ThePierreAuger:2015rma,AbuZayyad:2012ru}. 
The data of energy spectra and shower depth ($X_\text{max}$) distribution from surface detectors (SD)~\cite{Aab:2015bza} and fluorescence detectors (FD)~\cite{Aab:2014kda} have been measured with increasing precision. Analyses on the depth of the shower maximum distribution infer the nuclear mass of cosmic rays before they penetrate the atmosphere of the Earth. Besides, the arrival directions of the detected events are nearly isotropic, and the latest anisotropy data support their extragalactic origin~\cite{Aab:2017tyv,ThePierreAuger:2014nja,diMatteo:2020dlo}.

The observed cosmic-ray energy spectrum changes its qualitative behavior as a function of particle energy $E$. It steepens around $3\times 10^{15}$ eV (the so-called ``knee") and flattens around $3\times10^{18}$ eV (the ``ankle")~\cite{Lawrence:1991cc,Nagano:1991jz,Bird:1993yi}, and there is a strong suppression at the energy around $5\times10^{19}$ eV~\cite{Abbasi:2007sv,Abraham:2008ru}. 
The latter two features are meaningful to explain the physical properties of UHECRs. The formation of the ``ankle" can be interpreted naturally by the transition from Galactic to extragalactic sources (see reviews~\citep{Kotera:2011cp, Anchordoqui:2018qom,AlvesBatista:2019tlv}), \bz{or could be explained by the pair-production dip if the composition is dominated by protons~\cite{Berezinsky:2002nc,Aloisio:2006wv}}. 
The sharp decline of the UHECR flux beyond $5\times10^{19}$eV is consistent with the prediction of the Greisen-Zatsepin-Kuzmin (GZK) cutoff~\cite{Greisen:1966jv,Zatsepin:1966jv} or the similar cutoff due to the photodisintegration of heavy nuclei~\cite{Stecker:1969fw,Khan:2004nd,Allard:2008gj}, where UHECRs lose their energies by interactions with the cosmic background radiation during their propagation in intergalactic space. 
The flux suppression can also be caused by the limited maximum acceleration energy inside the sources~(e.g., Refs.~\cite{Aloisio:2009sj,Aab:2016zth}). 

The mass composition of UHECRs is crucial for us to test the theories concerning the origin, nature and production site of these particles~\cite{Abraham:2010yv,Abbasi:2014sfa}. 
Before finally being detected by the ground experiments, cosmic rays penetrate the atmosphere of the Earth and their hadronic interactions with the atmosphere produce extensive air showers (EAS) of secondary particles.
The EAS experiments often present the distributions of the mean depth of shower maximum $\langle X_{\rm max}\rangle$ and the Root Mean Square (RMS) $\sigma(X_{\rm max})$ of the $X_\text{max}$ as a function of energy to infer the composition of UHECRs~\cite{Aab:2014kda, Aab:2014aea}. The TA collaboration~\cite{Abbasi:2014sfa} also accumulated these events on the ground, and their processed results are consistent with the Auger results within the uncertainties~\cite{Abbasi:2015czo,Aab:2019ogu}, although the interpretations of the data are still under debate. 

Various efforts have been done in order to unveil the mysteries of UHECRs~\cite{AlvesBatista:2019tlv}, and an important piece of information is energetics. For any source class to be responsible for the observed UHECRs, their luminosity density has to be sufficiently larger than the energy generation rate density of UHECRs. 
In particular, the recent multimessenger observations have revealed that the UHECR generation rate density is comparable to the energy generation densities of PeV neutrinos and sub-TeV gamma rays, which may indicate physical connections among three multimessenger particles~\cite{Murase:2013rfa,Katz:2013ooa,Murase:2016gly,Fang:2017zjf,Murase:2018utn}. 

In this work, we revisit the energetics of UHECRs in light of the recent Auger data that support a nucleus-rich composition. The UHECR generation rate density has been studied especially for protons~\cite{Waxman:1995dg,Berezinsky:2002nc,Aloisio:2006wv,Katz:2008xx,Decerprit:2011qe}. 
There were less studies in light of the Auger data mainly for nucleus-rich composition models~\cite{Unger:2015laa,Aab:2016zth,Zhang:2017moz,Jiang:2020sgy}. 
We simulate the propagation of UHECRs with the public code \textsc{CRPropa 3}~\cite{Batista:2016yrx} for different source parameters and fitting energy ranges, and perform detailed spectral fits for different species of nuclei. Because it is not trivial how much the UHECR energetics is affected by different assumptions on the injection of nuclei, our results will be useful for modeling UHECR accelerators. 
We also utilize $X_\text{max}$ observations as additional constraints. As noted above, the interpretations of the $X_\text{max}$ data are under debate although the current data are consistent with mixed composition scenarios. Because a cosmic-ray energy scale corresponding to the energy scale reached by the Large Hadron Collider is still $\sim{10}^{17}$~eV, the extrapolation relying on uncertain hadronic interaction models are unavoidable in analyses of the observed EASs~\cite{Anchordoqui:2018qom}. In this sense, our results independent of the $X_\text{max}$ data are conservative and we show that all mixed composition results lie within the range of our obtained UHECR generation rate density.

This paper is organized as follows. In Sec.~II, we introduce the method and the model that we used to find the energy generation rate density and best-fit parameters to the spectral and composition data, as well as details of the setup of simulations. In Sec.~III, we present the results of spectral fits for different species of nuclei. In Sec.~IV, spectral and composition combined fits for these nuclei are shown. Finally in Sec.~V, we conclude and discuss implications of our results. 
In this work, we assume the $\Lambda$CDM cosmology with $\Omega_m=0.3$, $\Omega_\Lambda=0.7$, $\Omega_k = 0$ and $H_0$ = 70 km s$^{-1}$ Mpc$^{-1}$~\cite{Aghanim:2018eyx}.

\section{Method}
\subsection{Spectral fits}
We here consider ``bottom-up" scenarios for the UHECR sources, in which particles are accelerated by electromagnetic processes. 
Many acceleration mechanisms have been studied previously, such as shock acceleration (see Refs.~\cite{Blandford:1987pw,Drury:1983zz} for a review), shear/one-shot acceleration~\cite{Kimura:2017ubz,Mbarek:2019glq}, 
magnetic reconnection~\cite{Lovelace:1997,Giannios:2009}, and plasma wakefield acceleration~\cite{Chen:2002nd,Murase:2009pg,Ebisuzaki:2020ase}. 
We adopt a power-law source spectrum, with an exponential cutoff at the highest energies. This is a reasonable consequence of many acceleration models (but see also Ref.~\cite{Kimura:2017ubz}). 
The canonical power-law index expected in the theory of shock acceleration is $s\sim2$~\cite{Achterberg:2001rx,Kirk:2000yh,Kato:2000hd}, even though they may have a wide range depending on details of the shock Lorentz factor~\cite{Keshet:2004ch}, magnetization~\cite{Sironi:2010rb}, obliquity~\cite{Spitkovsky:2006np,Sironi:2009jw}, and escape processes~\cite{Ohira:2009rd}. 
We model the maximal acceleration energy of UHECRs at the sources are proportional to their charge $\propto Z$. This may be oversimplified in certain aspects, because interactions during the UHECR escape and propagation inside sources may vary cutoff energies among different species of nuclei, and the overall effective spectra with the summation of spectra from different sources~\cite{Unger:2015laa,Kotera:2015pya,Blaksley:2011kw,Fang:2017zjf}. 

This work focuses on one dimensional propagation, where the effect of large-scale extragalactic magnetic fields and Galactic magnetic fields are neglected, although this can in principle affect best-fit spectral indices of nuclei~\cite{Lemoine:2004uw,Kotera:2007ca,Mollerach:2013dza,Wittkowski:2017okb,Fang:2017zjf}. \add{See a brief discussion on the impact of intergalactic magnetic fields (IGMFs) in the Sec.~V.}
We assume that the sources of extragalactic UHECRs are distributed along the line-of-sight with minimum redshift $z_{\rm min}=0.0001$ and maximum redshift $z_{\rm max}= 2.1$. We use simulations to verify that the particles from $z>2.1$ only contribute less than 1\% of UHECRs arrived at Earth due to the interactions during propagation discussed below.

The observed flux of UHECR nuclei with mass number $A$ is calculated using the following formula,
\begin{eqnarray}
\Phi_A(E) &=& \sum_{A^\prime} \frac{c}{4\pi} \int_{z_{\rm min}}^{z_{\rm max}} dz  \left| \frac{dt}{dz}\right| F(z) \nonumber \\ &\times & \int_{E^\prime_{\rm min}}^{E^\prime_{\rm max}} dE^\prime \frac{d\dot{N}_{A^\prime}}{dE^\prime}\frac{d\eta_{A A^\prime}(E, E^\prime,z)}{dE},\,\,\,\,\,\,\,\,\,\,
\end{eqnarray}
where \bz{$d\dot{N}_{A^\prime}/dE$ is the differential injection rate of UHECR nuclei injection rate per unit volume, and} 
\begin{eqnarray}
\frac{dt}{dz} = -\frac{1}{H_0(1+z)}\frac{1}{\sqrt{\Omega_\Lambda+\Omega_k(1+z)^2+\Omega_m(1+z)^3}},
\end{eqnarray}
and $F(z)=(1+z)^m$ is assumed to implement the redshift evolution of the luminosity density of the UHECRs for the redshift evolution index $m$, and $\eta_{A A^\prime}(E, E^\prime,z)$ is the fraction of generated cosmic rays with mass number $A$ and energy $E$ from parent particles with mass number $A^\prime$ and energy $E^\prime$~\cite{Zhang:2017hom}. 

For the injection from the sources, we assume a power-law distribution of UHECRs from identical sources with one species of nuclear injection,
\begin{equation}
{E'}^2 J_A^{\rm inj}(E')\equiv 
{E'}^2\frac{d\dot{N}_{A}}{dE'} =C_A^{\rm inj}{\left(\frac{E'}{E'_0}\right)}^{2-s}\text{exp}\left(-\frac{E'}{E'_\text{max}}\right)
\end{equation}
where $E'$ is the energy of injected nuclei, $s$ is the spectral index, $E'_{\rm max}$ is the maximum energy, and $E'_0(\ll E'_{\rm max})$ is the reference energy at which the normalization factor $C^{\rm inj}$ is defined.
We only consider certain species of primary nuclei injected at the sources, for example proton, helium, oxygen, silicon and iron, but secondary nuclei are generated during the propagation via photonuclear interactions.

When UHECRs propagate in intergalactic space, they will undergo various energy loss processes via interactions with ambient photon backgrounds, including the photomeson production, Bethe-Heitler pair production, and photodisintegration processes~\cite{Batista:2016yrx}.
\textsc{CRPropa 3} takes an advantage of external packages, such as \textsc{TALYS}~\cite{etde_21285928} and \textsc{SOPHIA}~\cite{Mucke:1999yb}, to conduct the Monte-Carlo simulations of nuclear reactions and photohadronic processes beyond the resonance range, and adiabatic energy losses are taken into consideration as well.
The ambient photon backgrounds consist of the cosmic microwave background (CMB) and the extragalactic background light (EBL) that are mainly the cosmic optical and infrared Backgrounds. We use the EBL model provided in Ref.~\cite{Gilmore:2011ks}. 

In this work, we consider $-2\leq m\leq 7$. In general, softer spectra, i.e., larger spectral indices can be compensated by smaller values of $m$ (e.g., Ref.~\cite{Takami:2007pp}). 
Throughout we scan the full range of $m$, but one should keep in mind the importance of neutrino and gamma-ray constraints on possible values of $m$. 
First, extremely-high-energy cosmic neutrino data from IceCube has ruled out source evolution models that are faster than the star formation rate if UHECRs is proton-dominated~\cite{Aartsen:2018vtx}. 
The diffuse neutrino fluxes from UHECR nuclei are predicted to be lower by an order of the magnitude compared to the ones from the proton-dominated cases~\cite{Murase:2010gj} for the same $m$. The constraints from the IceCube data on $m$ are expected to be as weak as $m\lesssim6$, especially if the nucleus-survival condition is satisfied.
\km{Second, strong redshift evolution models with $m\gtrsim5$ are independently at variance with the existing gamma-ray constraints from {\it Fermi}-LAT measurements of the diffuse isotropic gamma-ray background~\cite{Berezinsky:2010xa,Supanitsky:2016gke}. 
While these ``multimessenger'' constraints are important, this work focuses on results obtained by the UHECR measurements, so it is fair to a wider range of evolution parameters $-2<m<7$ given that there are degeneracies with the spectral index and composition. 
}
Even if the allowed values of $m$ are narrowed down to $m\lesssim3$, the results on the energetics are not much affected, as shown in the next section.

In order to find the best-fit spectrum to the observed UHECR data, we scan over several parameters that are the spectral index $s$, maximum UHECR energy $E_\text{max}$, and redshift evolution index $m$.
The systematic uncertainty in the UHECR energy measurements is of the order of 10\%~\cite{Aab:2017njo}. 
As a result of this possible energy shift, the reconstructed events may have inaccurate energy and be classified into the false energy bin in the spectrum, which affects the measured flux. To take this into consideration and find the expected event counts in each of the energy bins (labelled by integer $i$), we introduce a free parameter $\delta_E$ as $\hat{E}_i\equiv (1+\delta_E)E_i$ to the simulated spectrum~\cite{Heinze:2015hhp}, which is combined with the other three source parameters ($s$, $E_\text{max}$ and $m$) to be searched for. Then we estimate the goodness of fitting to the observed energy spectrum using the chi-square method~\cite{Heinze:2015hhp,Fang:2017zjf}.
\begin{eqnarray}
\chi^2_\text{spec}=&&\sum_i \frac{(f \Phi^\text{sim}(\hat{E}_i; s, E_\text{max}, m)-\Phi^\text{Auger}(E_i))^2}{\sigma_i^2}\nonumber\\ &&+\left(\frac{\delta_E}{\sigma_E}\right)^2,
\end{eqnarray}
where $f$ is the free normalization factor induced from the UHECR energy generation rate density, $\Phi^\text{sim}(\hat{E}_i; s, E_\text{max}, m)$ is the simulated unnormalized flux at Earth from our model evaluated at $\hat{E}_i$, and $\Phi^\text{Auger}(E_i)$ is the UHECR flux measured by Auger. 
For given $E_i$, $\sigma_i$ contains both systematic and statistic errors in the flux data, and the systematic uncertainty of the measured energy scale is $\sigma_E=14\%$~\cite{Aab:2017njo}. The range of $\delta_E$ is from $-14\%$ to $14\%$ and the statistic uncertainty considered in the fitting procedure is $5\%$. Dividing $\chi^2$ by the degree of freedom (d.o.f.) has often been used as a statistical tool in the hypothesis test to examine whether a model can be ruled out for a given fit. The minimal $\chi^2$ is denoted as $\chi_{\rm min}^2$ throughout this work. 

The free normalization factor ($f$) that is determined by the spectral fits is directly used for the determination of the UHECR energy generation rate density, $EQ_{E}(z)=F(z)({E}^2 J^{\rm inj}$).
In general, the ``total'' UHECR luminosity density (or energy generation rate density) at $z=0$ is calculated as
\begin{equation}
Q =\int_{E_{\rm min}} dE \, Q_{E}= \sum_{A} \int_{E_{\rm min}} dE\, E\frac{d\dot{N}_A}{dE},
\label{totalbudget}
\end{equation}
where $E_{\rm min} = A m_p c^2$ is the minimum CR energy. 
The differential UHECR energy generation rate density at the reference energy $E_r$ is given by
\begin{equation}
E Q_{E}^{\log_{10}(E_r)}={E}^2 J^{\rm inj}|_{E_r}=\sum_{A}\left.{E^2\frac{d\dot{N}_A}{dE}} \right\rvert_{E_r},
\end{equation}
where $E_r=10^{19.5}$~eV is adopted as a fiducial reference energy. \km{Note that we mainly use the differential UHECR energy generation rate density as a proxy. This is because the total cosmic-ray energy generation rate density is highly model dependent and the value varies depending on the literature. It is sensitive to not only the power-law index but also the integral threshold. The total value is still useful when it is compared to energetics of the sources~\cite{Murase:2018utn}, but it can always be derived from the differential one through the integration.}
Using $\Phi^{\rm Auger}(E)$ as well as the simulated $\Phi^{\rm sim}(\hat{E}_i)$, we can determine the value of $C_A^\text{inj}$ through $f$. Scanning over $f$ (or $C_A^\text{inj}$) to fit the Auger spectrum by simulated spectra in some energy range, we find the best-fit ($\chi_{\rm min}^2$) parameters ($s$, log$_{10}(E_\text{max})$, $m$), and determine the corresponding UHECR energy generation rate density $EQ_E^{19.5}$.

\subsection{Composition fits}
There is another observable, $X_\text{max}$, which gives information on the composition of the observed UHECRs. When UHECRs arrive at Earth, they will interact with the atmosphere and generate EASs. FDs can measure the energy profile of the shower and the depth corresponding to the maximum development of the shower, $X_\text{max}$, which is fundamental to determine the nature of the primary cosmic rays that initiate cascades. SDs can also measure the energy of primary particles at ground level with a full duty cycle~\citep{Aab:2019ogu}. 

If we are aware of mass number $A$ and energy $E$ of primary particles, we are able to estimate the distribution of $X_\text{max}$. However, one has to rely on hadronic interaction models involved in the shower development, such as EPOS-LHC~\cite{Pierog:2013ria}, Sybill 2.1~\cite{Ahn:2009wx} and QGSJet II-04~\cite{Ostapchenko:2010vb}. Although there are still significant uncertainties in the interaction models, the mass composition before penetrating Earth atmosphere can be directly transformed into consequential $X_\text{max}$ distribution. 

For given hadronic interaction models, by fitting both the spectrum and $X_\text{max}$ distribution parameters (whose first two statistical moments are $\langle X_\text{max} \rangle$ and $\sigma^2(X_\text{max})$), one can also constrain source composition models~\cite{Kampert:2012mx,AlvesBatista:2019tlv,Taylor:2011ta,Heinze:2019jou,Zhang:2017moz}. 
We do not simulate realistic air showers or cascade processes. Instead, we follow the phenomenological method, which enables us to calculate $X_\text{max}$ properties based on the percentage information of different nuclei arriving at Earth in our simulation. 
The $X_\text{max}$ distribution is a function of particle energy and mass of nuclei entering at the Earth's atmosphere. 
We use the generalized Gumbel distribution function~\cite{DeDomenico:2013wwa} to describe the underlying probability distribution of $X_\text{max}$, which is defined as 
\begin{equation}
\mathcal{G}(z)= \frac{1}{\sigma}\frac{\lambda^{\lambda}}{\Gamma(\lambda)}(e^{-\lambda z-\lambda e^{-z}}),\qquad z= \frac{x-\mu}{\sigma},
\end{equation}
where $X_\text{max}$ is taken as $x$ in this distribution function, and the other parameters $\mu$, $\sigma$ and $\lambda$ are obtained by the following equations sets: $\mu(A,E)= p_{0_{\mu}} + p_{1_{\mu}} \log_{10} (E/E_{0})+ p_{2_{\mu}} \log_{10}^{2} (E/E_{0})$, $\sigma(A,E)= p_{0_{\sigma}} + p_{1_{\sigma}} \log_{10} (E/E_{0})$, and $\lambda(A,E)= p_{0_{\lambda}} + p_{1_{\lambda}} \log_{10} (E/E_{0})$. Here $E$ and $A$ are the energy and mass of the primary particle, respectively, and $E_{0}= 10^{19}$~eV is a reference energy. The $A$ dependence of the parameters are empirically found as: 
$p_{0}^{\mu,\sigma,\lambda}= a_{0}^{\mu,\sigma,\lambda} + a_{1}^{\mu,\sigma,\lambda}\ln A + a_{2}^{\mu,\sigma,\lambda}\ln^{2} A$, $p_{1}^{\mu,\sigma,\lambda}= b_{0}^{\mu,\sigma,\lambda} + b_{1}^{\mu,\sigma,\lambda}\ln A + b_{2}^{\mu,\sigma,\lambda}\ln^{2} A$, $p_{2}^{\mu}= c_{0}^{\mu} + c_{1}^{\mu}\ln A + c_{2}^{\mu}\ln^{2} A$.
The parameters $a$, $b$, $c$ are obtained from the CONEX shower simulation, which is a Monte Carlo simulation of high-energy interactions using a numerical solver to the cascade equations to calculate distributions of secondary particles, based on different hadronic interaction models for different nuclei~\cite{Pierog:2004re}. Because the collider experiments on Earth have not been able to generate energy up to several hundreds of TeV in the center-of-momentum frame, the parameters are extrapolated from lower-energy experimental data, which brings significant uncertainties in these analysis. 
The values of $\mu$ and $\sigma$ are related to the mean and the variance of the underlying distribution of $X_\text{max}$, respectively. 
Because the direct simulations of all showers generated by primary particles is time consuming, the use of the generalized Gumbel distribution is more efficient to calculate the total $X_\text{max}$ distribution under different compositions and it is sufficient for the purpose of this work (see Ref.~\cite{DeDomenico:2013wwa} for details, where three hadronic interaction models, EPOS-LHC, QGS Jet II-04, and Sybill 2.1, are considered). In our calculations, nuclei with the mass number $A$ between 1 and 56 are counted.

The probability density function (PDF) of $X_\text{max}$ would depend on $E$ and $A$. We have different species of nuclei arriving at Earth with different energies. We need to sum up the results for different species with a weight of their percentage in each energy bin. This leads to a final simulated $X_\text{max}$ distribution outcome that can be compared to the observables, namely $\langle X_\text{max} \rangle$ and $\sigma(X_\text{max})$, from the experiments~\cite{Aab:2014aea, Aab:2014kda, Abraham:2010yv, Abbasi:2014sfa}. 

As in the spectral fits described above, we use the chi-square method calculated by the following equation to estimate the goodness of $\langle X_\text{max} \rangle$ and $\sigma(X_\text{max})$ fit, e.g.,
\begin{eqnarray}
\chi^2_{\langle X_\text{max} \rangle} =&& \sum_i \frac{(\langle X_\text{max}^\text{sim}\rangle(\hat{E}_i; s, E_\text{max}, m)-\langle X_\text{max}^\text{Auger}\rangle(E_i))^2}{\sigma_i^2}\nonumber\\
&&+ \left(\frac{\delta_E}{\sigma_E}\right)^2.
\end{eqnarray}
After the acquisition of the $\chi^2$ distributions for the three observables (namely the spectrum, $\langle X_\text{max} \rangle$ and $\sigma(X_\text{max})$), we obtain the total $\chi^2/\rm d.o.f$ of the combined fits,
\begin{equation}
\frac{\chi^2}{\rm d.o.f} = \frac{\chi^2_\text{spec}+\chi^2_{\langle X_\text{max} \rangle}+\chi^2_{\sigma(X_\text{max})}}{\rm d.o.f_\text{spec}+\rm d.o.f_{\langle X_\text{max} \rangle}+\rm d.o.f_{\sigma(X_\text{max})}},
\end{equation}
which can be used to evaluate the overall goodness of fit, considering both the spectrum and mass composition of the UHECRs. If the full information on the $X_\text{max}$ distributions rather than their first two statistical moments is taken into account for fitting, the derivation would be more accurate and the possible coincidence of $\langle X_\text{max} \rangle$ and $\sigma(X_\text{max})$ out of different distributions~\cite{Aab:2014aea} would be totally avoided. Thus it is not surprising that our results slightly differ from those of Ref.~\cite{Aab:2016zth}.

To determine the fitting parameters, we perform uniform scans over three dimensional grids of the spectral index $s$, the maximum energy $E_{\rm max}$, and redshift evolution index $m$. On each grid point, the discrepancy between simulated and observed data is a function of the UHECR energy generation rate density. We find the best-fit parameters, at which the UHECR energy generation rate density is evaluated. 
We consider $s$ from 0.0 to 2.9 with an interval of 0.1, covering the typical range predicted by the shock acceleration mechanism ($s\sim 2.0 - 2.2$). The range of $E_{\rm max}$ is from $10^{19.0}$eV to $10^{21.0}$eV and the range of $m$ is from -2 to 7 with an interval of 1.

\section{Results of spectral fits}
We adopt a power-law injection spectrum and fix the EBL and hadronic interaction models in order to focus on the parameters of our interest. The propagation simulation is completed by \textsc{CRPropa 3} with the Gilmore EBL model~\cite{Gilmore:2011ks}.

The experimental data used for the fits include two parts: the event distribution in energy bins of 0.1 in $\log_{10}(E/\text{eV})$, and $X_\text{max}$ distribution in the exact same energy bins but up to $\log_{10}(E/\text{eV})=19.5$ and a final bin combining data $\log_{10}(E/\text{eV})$ from 19.5 to 20.0, whose average energy $\langle \log_{10}(E/\text{eV})\rangle=19.62$. For the purpose of comparison and fitting, we divide our simulated events into exact same energy bins.

We consider two fitting ranges from $\log_{10}(E/\text{eV})=18.45$ to $\log_{10}(E/\text{eV})=20.15$ and from $\log_{10}(E/\text{eV})=19.05$ to $\log_{10}(E/\text{eV})=20.15$, respectively. In general, the choice of the fitting range depends on scenarios of the transition from Galactic to extragalactic components. 
We consider the first case as the fiducial energy range because the UHECR spectrum below $\sim{10}^{19.1}$~eV shows a good agreement within systematic errors. However, one should keep in mind that the second case can give more conservative estimates on $EQ_E$ by a factor of 2 (see below).

\subsection{Proton}
\begin{figure*}[tb]%
\centering
\subfloat[Best-fit spectrum for proton]{{\includegraphics[width=0.33\linewidth]{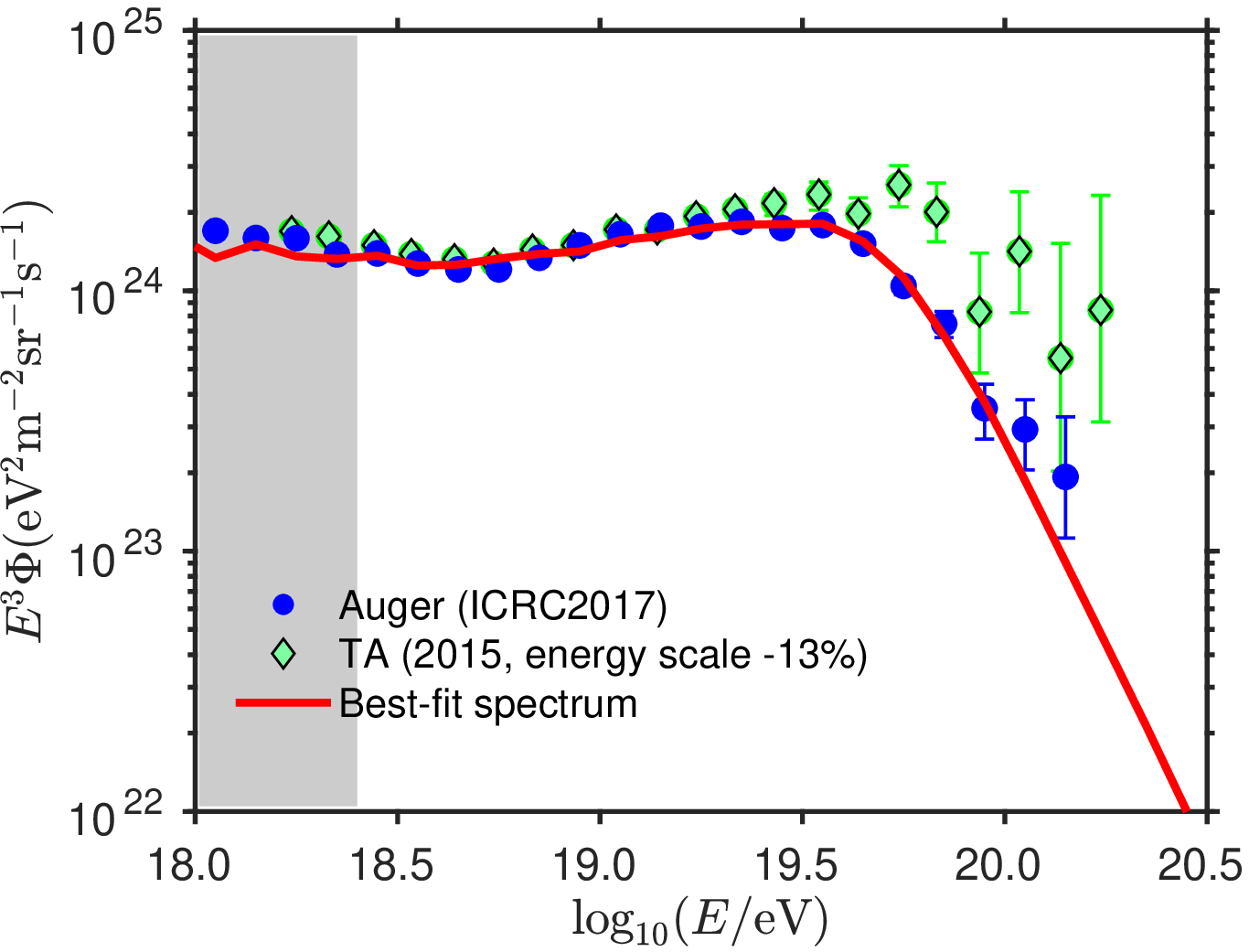} }}%
\subfloat[$\chi_{\rm min}^2$/d.o.f. as a function of $s$ and $m$, where $E_\text{max}$ is scanned.]{{\includegraphics[width=0.33\linewidth]{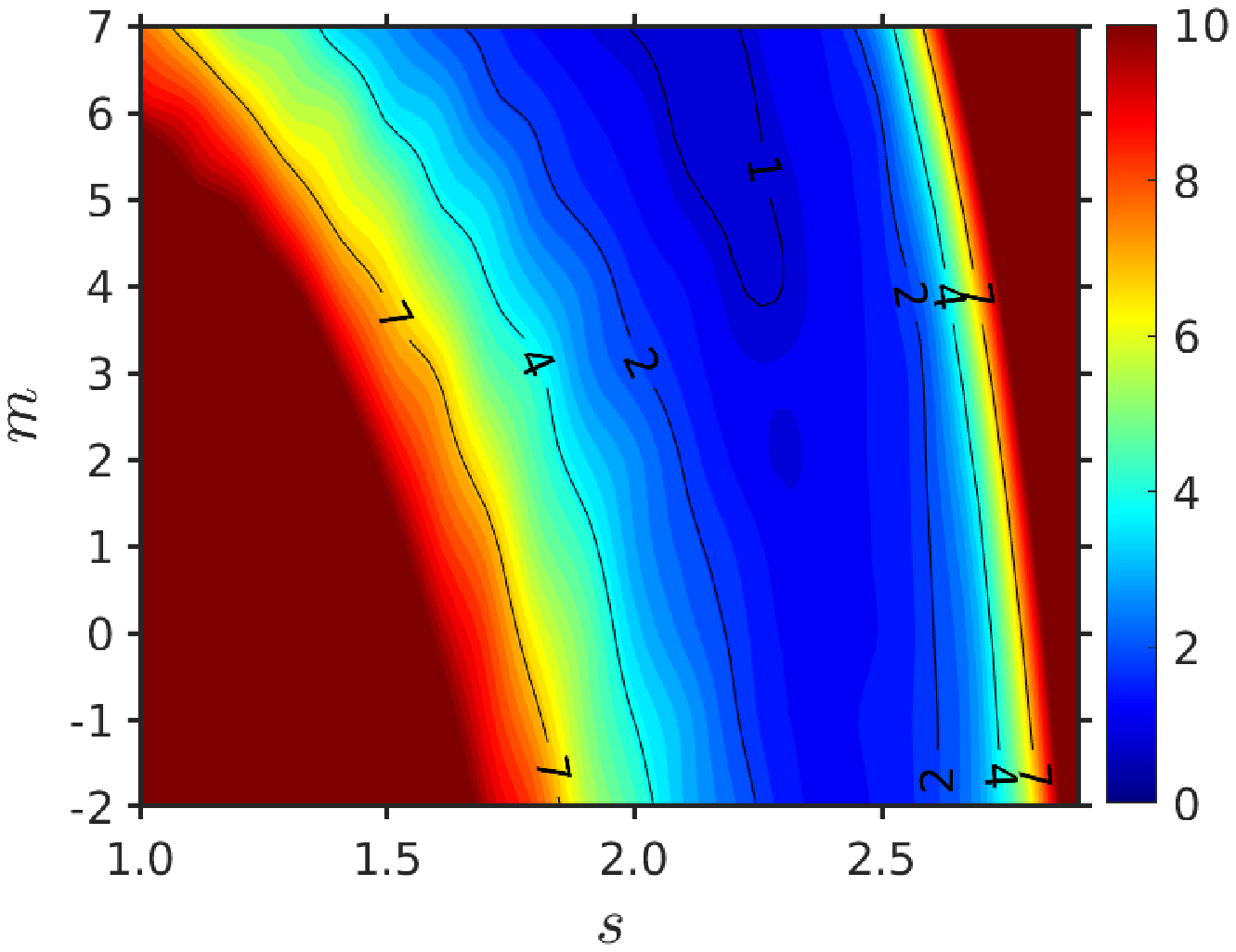} }}%
\subfloat[$\chi_{\rm min}^2$/d.o.f. as a function of $s$ and $E_\text{max}$, where $m$ is scanned.]{{\includegraphics[width=0.33\linewidth]{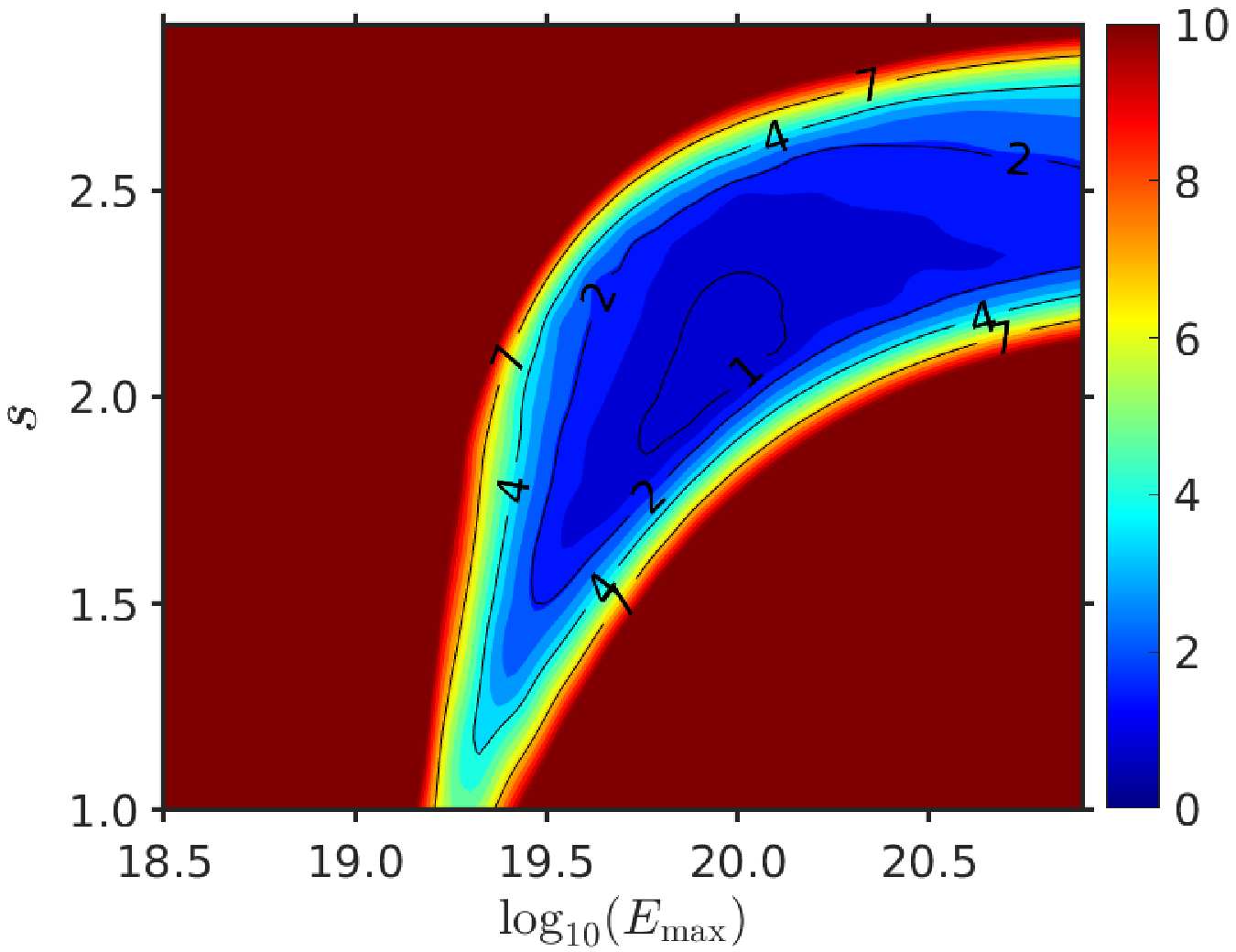} }}%
\caption{
(Left) Our best-fit energy spectrum for the Auger data in the pure proton composition case. The differential simulated UHECR flux ($\Phi$) before entering the terrestrial atmosphere is multiplied by $E^3$. The best-fit parameters are $E_{p,\text{max}} = 10^{19.9}$~eV, $s = 2.1$, $m=5.0$ and $\delta_E =0.06$. The shaded region indicates where the data is not included in the fit. 
(Middle) Best-fit parameter space for the spectral index $s$ and redshift evolution index $m$ in the pure proton composition case. The solid contours indicate certain values of $\chi_{\rm min}^2$/d.o.f., where smaller than 2 indicate legitimate fits to the spectrum. 
(Right) Best-fit parameter space for $s$ and $E_\text{max}$.}%
\label{fig:3DscanProton}%
\end{figure*}

\begin{figure*}%
\centering
\subfloat[Fitting energy range from $10^{18.45}$~eV to $10^{20.15}$~eV]{{\includegraphics[width=0.5\linewidth]{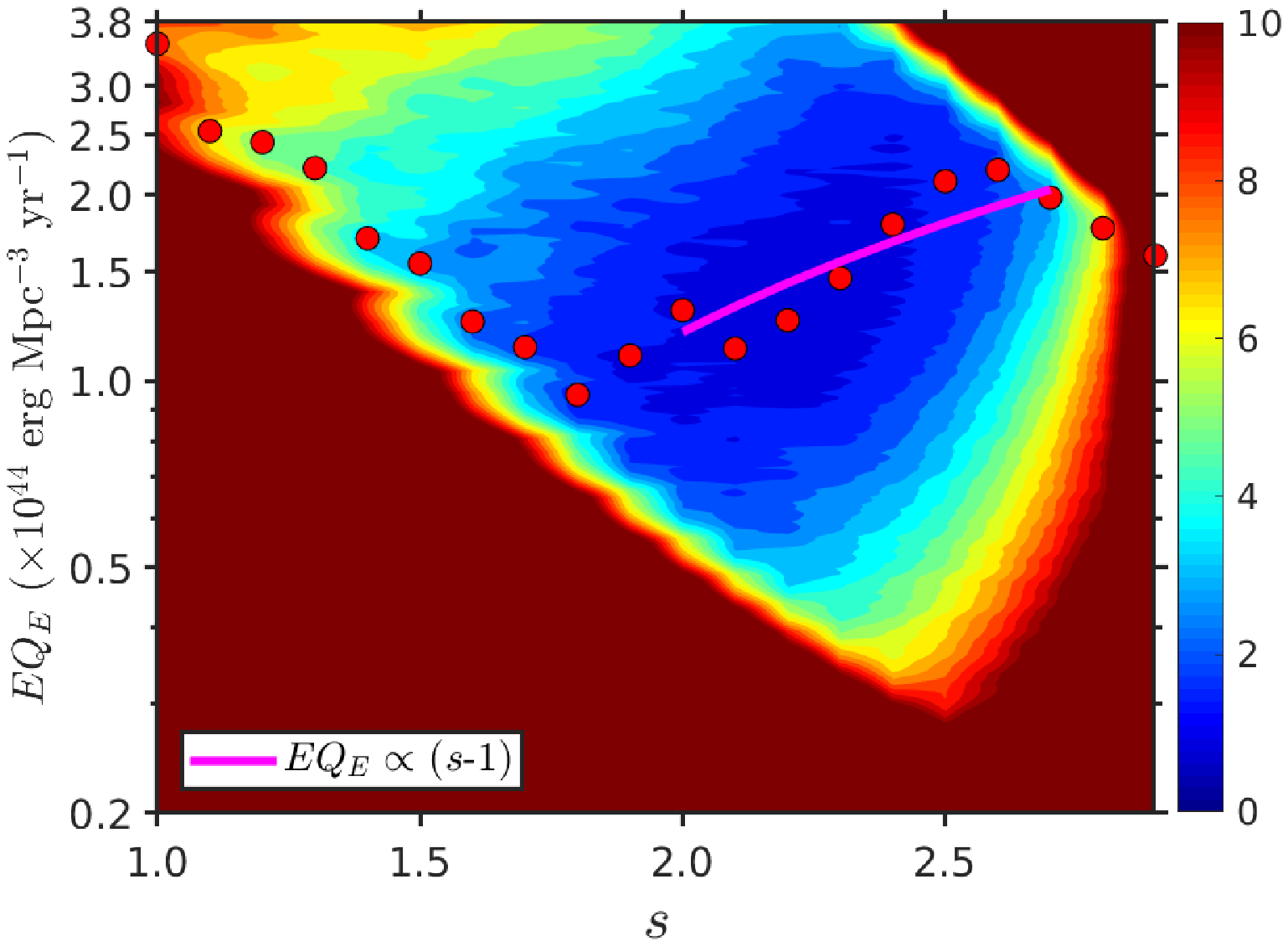} }}%
\subfloat[Fitting energy range from $10^{19.05}$~eV to $10^{20.15}$~eV]{{\includegraphics[width=0.5\linewidth]{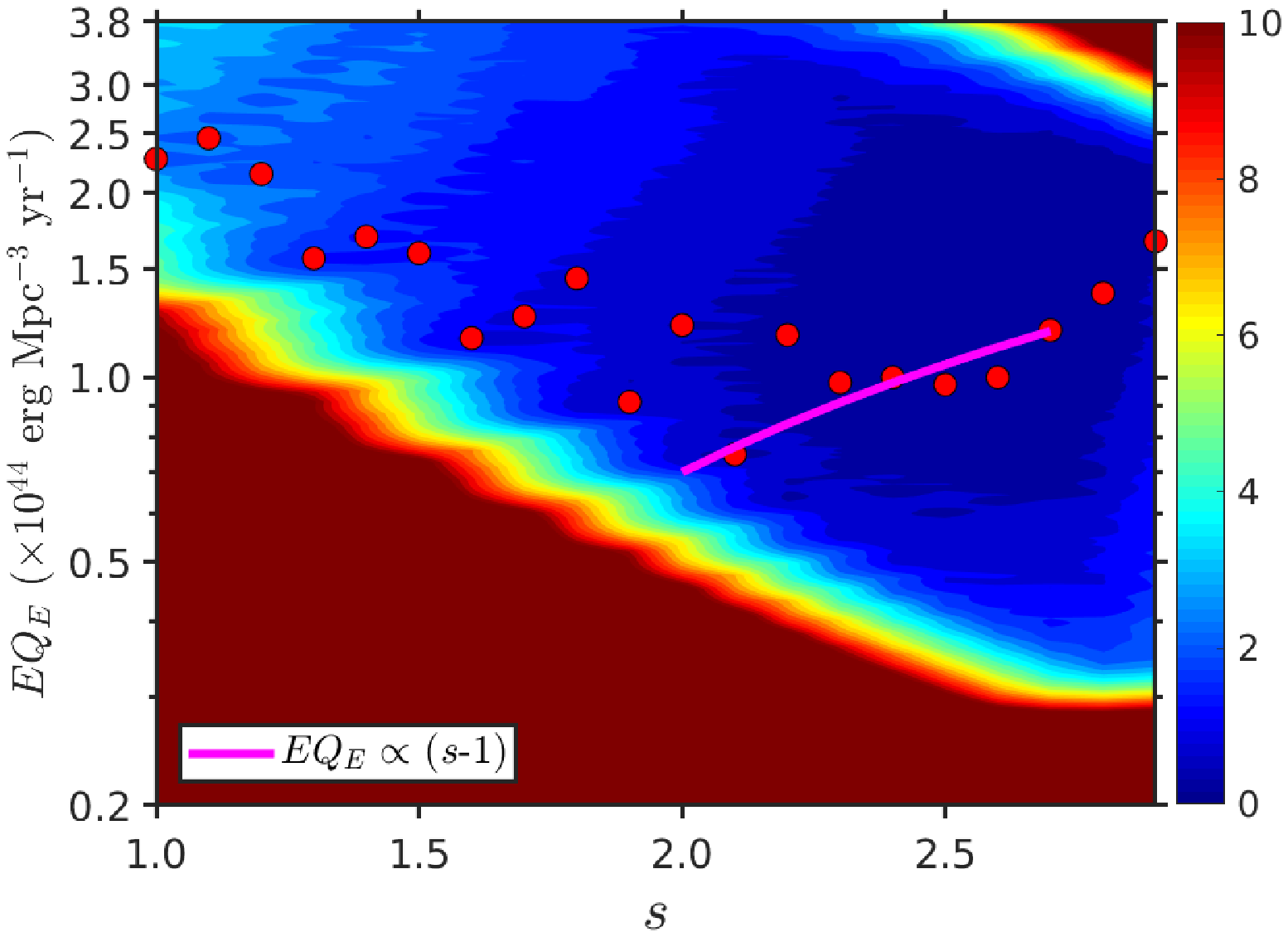} }}
\caption{Best-fit energy generation rate density $EQ_E^{19.5}$ as a function of the spectral index $s$ (red solid dots), compared with the analytic dependence of $EQ_E^{19.5} \propto s-1 $ in the $s$ range of [2.0, 2.7]. 
Two parameters, $m$ and $E_\text{max}$, are scanned. The background contour reflects the $\chi_{\rm min}^2/\rm d.o.f$ with different combinations of the energy generation rate density $EQ_E^{19.5}$ and power-law index $s$. Pure proton composition at the sources is assumed. 
The left and right plots correspond to the results from two different fitting energy ranges.}
\label{fig:injectionWithGammaProton}%
\end{figure*}

\begin{figure}
\includegraphics[width=1.0\linewidth]{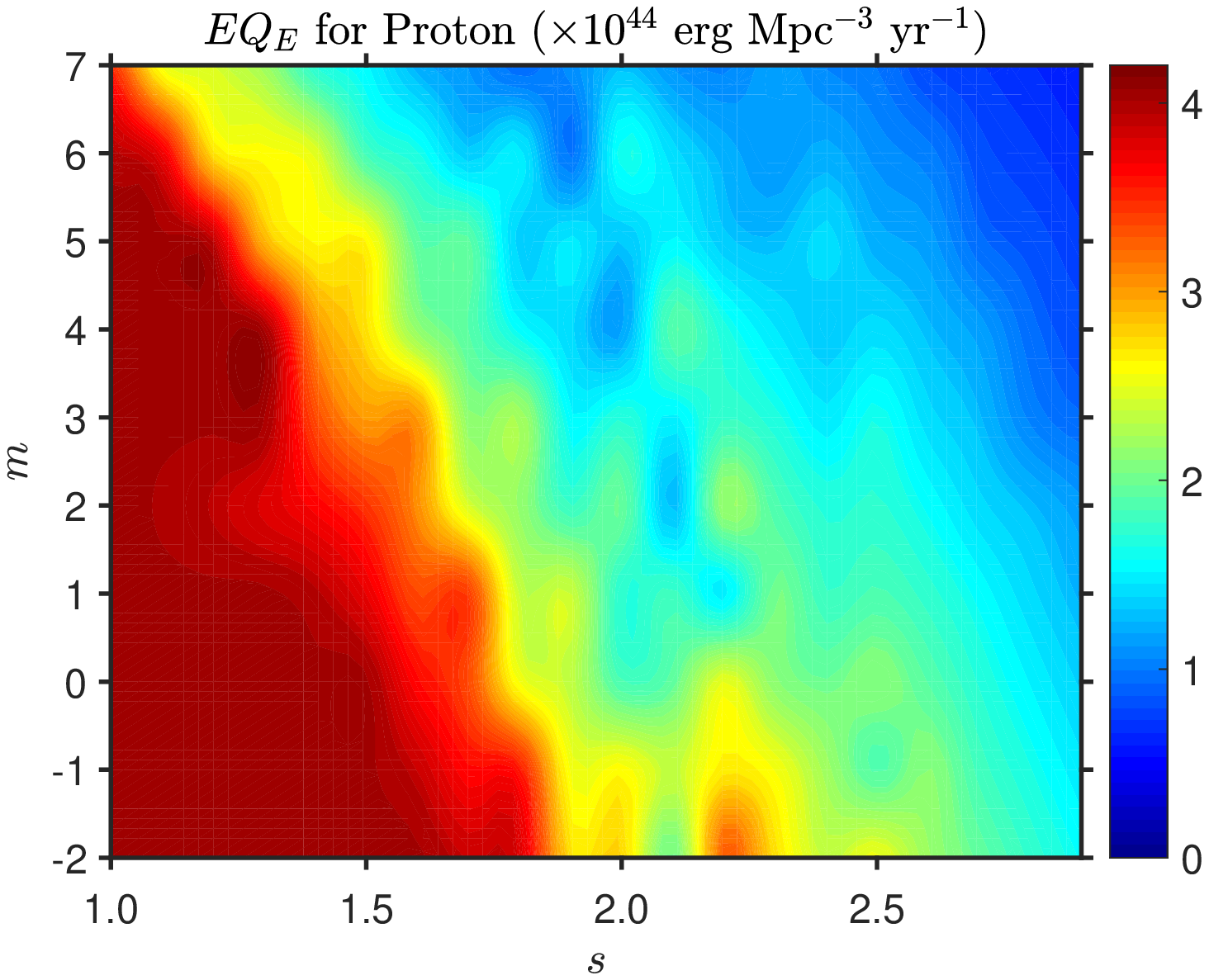}
\caption{Contour showing values of the UHECR generation rate density $EQ_E^{19.5}$ at $\chi_{\rm min}^2$ determined by $s$ and $m$, where $E_\text{max}$ is marginalized.
The fitting energy range is from $10^{18.45}$ eV to $10^{20.15}$ eV.
\label{fig:Proton3DInjection}}
\end{figure}

First, we show the results for the pure proton composition case. 
The fit for the Auger data from $10^{18.45}$ eV to $10^{20.15}$ eV is shown in Fig.~\ref{fig:3DscanProton}(a).
Note this fitting energy range is used for the following analysis, unless otherwise noted. 
The best-fit parameters in this setup are $E_{p,\text{max}} = 10^{19.9}$~eV, $s = 2.1$, $m=5.0$ and $\delta_E =0.06$. The chi-square value is $\chi^2/\rm d.o.f = 0.78$ which corresponding to a p-value of 71.98\%. 
The results are consistent with previous studies~\cite{Abraham:2009wk,Murase:2008sa,Katz:2008xx}.
For example in Ref.~\cite{Katz:2008xx}, they derived their parameters to be $E_{p,\text{max}} = 10^{19.6}$~eV, $s = 2.2$ and $m=3.0$.
Fig.~5 in Ref.~\cite{Abraham:2009wk} showed a fit with pure proton composition to the older Auger data using a parameter set similar to ours: $E_{p,\text{max}} = 10^{20.5}$~eV, $s = 2.3$ and $m=5.0$. 
Although we use the Auger data for the purpose of estimating the UHECR energy generation rate density, we also show the TA 2015 data (see Fig.~\ref{fig:3DscanProton}(a)) for comparison, with a change of the energy scale by 13\% to make it in line with the Auger spectrum around the ankle. 
Our results are also consistent with Ref.~\cite{Heinze:2015hhp}, although some differences exist. We fit the spectrum for the Auger 2017 data, while they used the TA 2015 data. Their value of best-fit index, $s\sim1.5- 1.6$, is smaller than the indices we get ($s\sim2.1-2.2$). 

In Fig.~\ref{fig:injectionWithGammaProton}, we show the contour plots which indicate the goodness of fit to the observed energy spectrum as a function of differential energy generation rate density $EQ_E^{19.5}$ and spectral index $s$. The red dots shown in the figure indicate the best-fit $EQ_E^{19.5}$ for given $s$. For $2\lesssim s\lesssim2.7$, the values of $EQ_E^{19.5}$ we get are similar with those of previous works.
For example, Refs.~\cite{Waxman:1995dg,Bahcall:2002wi,2008AdSpR..41.2071B,Murase:2008sa,Katz:2008xx} showed $Q(E>10^{19.5}{\rm~eV})\sim (0.3 - 1.0)\times10^{44}$ erg Mpc$^{-3}$ yr$^{-1}$.
Ref.~\cite{Katz:2008xx} also indicated $EQ_E^{19.5} \propto (s-1)$. 
However, this linear relation between $E Q_E^{19.5}$ and $s$ is not robust when we scan over other parameters like $m$ and $E_{\rm max}$.
The values of $EQ_E^{19.5}$ also tends to be larger for $s\lesssim2$. This is because more UHECRs are depleted for hard spectra so one needs to inject more UHECRs to match the data. 

We also find that the results on $EQ_E^{19.5}$ depend on the fitting energy range (see Appendix~\ref{appendix:fit} for the full discussion).
In the right plot in Fig.~\ref{fig:injectionWithGammaProton}, we show the results with an energy fitting range from $10^{19.05}\rm~eV$ to $10^{20.15}\rm~eV$. 
Statistically, fewer data points to be fit make smaller values of $\chi_{\rm min}^2/\rm d.o.f.$ possible. However, we see that the shape of the contour in Fig.~\ref{fig:injectionWithGammaProton} and the correlations between parameters, e.g. $s$ and $m$, remain similar when the fitting energy range changes (also see Appendix~\ref{appendix:fit}). 

In the proton case, the differential energy generation rate density $EQ_E^{19.5}$ is larger for harder spectra indices $s$, but very small values of $s$ are not preferred by the fit.
Middle and right plots in Fig.~\ref{fig:3DscanProton} present contours of $\chi_{\rm min}^2/\rm d.o.f.$ as a function of ($s$, $m$) and ($E_\text{max}$, $s$): the distribution of $\chi_{\rm min}^2/\rm d.o.f.$, corresponding to the best fits, behaves like valley curves in the figure. It is clear to see that the $\chi_{\rm min}^2$ region (or good fit region) spans widely in the range of $s$. For $2.0<s<2.5$, the change of $s$ does not strongly affect the spectra of the observed UHECRs and thus resulting $\chi_{\rm min}^2$ from fitting spectra. 
The reason might be that the value of $E_\text{max}$ is low enough to make accelerators near the $E_\text{max}$ be the dominant contributors to the observed UHECR spectra. Therefore, the change of $s$ becomes relatively unimportant in the spectrum on the Earth.
For $s < 2.0$, only high-redshift evolution $m > 5$ or lower maximum energy $E_\text{max}(<10^{20.0}\rm eV)$ can give a reasonable fit. There are certain regions of the parameters that can give reasonable good fits ($\chi^2/\rm d.o.f$ smaller than 2, corresponding to a p-value of 0.84\%), indicated by blue color in the contours.
Actually, these regions show clear correlations among the parameters we searched ($s$, $E_\text{max}$, $m$), and that is the case for different species of nuclei and their mixed injection as we show later. For instance, when the value of $m$ gets smaller, a higher value of $s$ is required in order to produce lower values of $\chi^2$. However, these general features are not exactly the same for different nuclei. For example, in the case of proton, there is one single minimum for the $\chi^2/\rm d.o.f$, while for some heavier nuclei, there are more than one local minimum, which means that there are multiple regions with good fits but are not adjacent to each other.

As shown in Fig.~\ref{fig:3DscanProton}, when the value of $E_\text{max}$ is large, the maximal energy at the source has little impact on the decline of the observed spectrum at high energies, and the interactions with cosmic background photons (dominated by the CMB in this energy) during propagation are the main cause. The soft spectrum also plays a role in this case. As a result, large values of $E_\text{max}$ accompanied by sufficiently large power-law indices ($s$) can lead to good spectral fits. With lower values of $E_\text{max}$, both the intrinsic maximum energy at the sources and the propagation cutoff lead to the flux suppression, which is consistent with the observed spectrum. Lower values of $\chi^2$ in the figure indicate a trend that lower values of $E_\text{max}$ correlate with smaller values of $s$. 

There are certain patterns and fluctuations in the parameter dependence of $E Q_E^{19.5}$ (see Fig.~\ref{fig:Proton3DInjection}.
The value of $EQ_E^{19.5}$ is larger for smaller values of $s$, but too small indices do not give good fits because a very hard spectrum requires a very low value of the maximum energy to have a cutoff that is consistent with the data (see Fig.~\ref{fig:3DscanProton}(b)).
On the other hand, the value of $EQ_E^{19.5}$ is smaller for larger $s$ and larger $m$. For a given $s$, stronger evolution models gives smaller values of $EQ_E^{19.5}$, which is consistent with our result in Fig.~\ref{fig:Proton3DInjection}.

\subsection{Light and intermediate nuclei}
\begin{figure*}
\centering
\subfloat[Spectral fit for He]{{\includegraphics[width=0.5\linewidth,trim={0 0cm 0 0},clip]{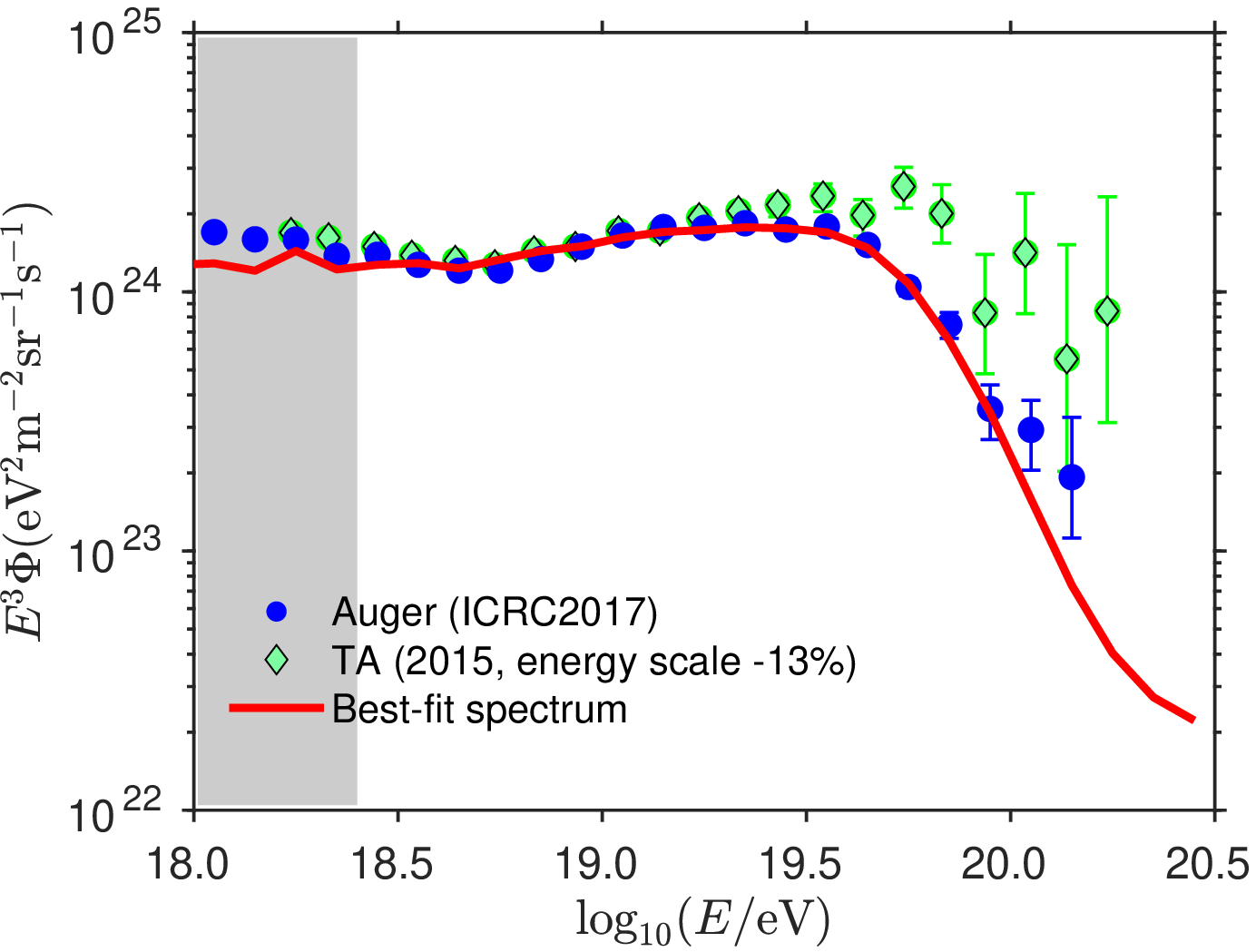} }}
\subfloat[Spectral fit for O]{{\includegraphics[width=0.5\linewidth]{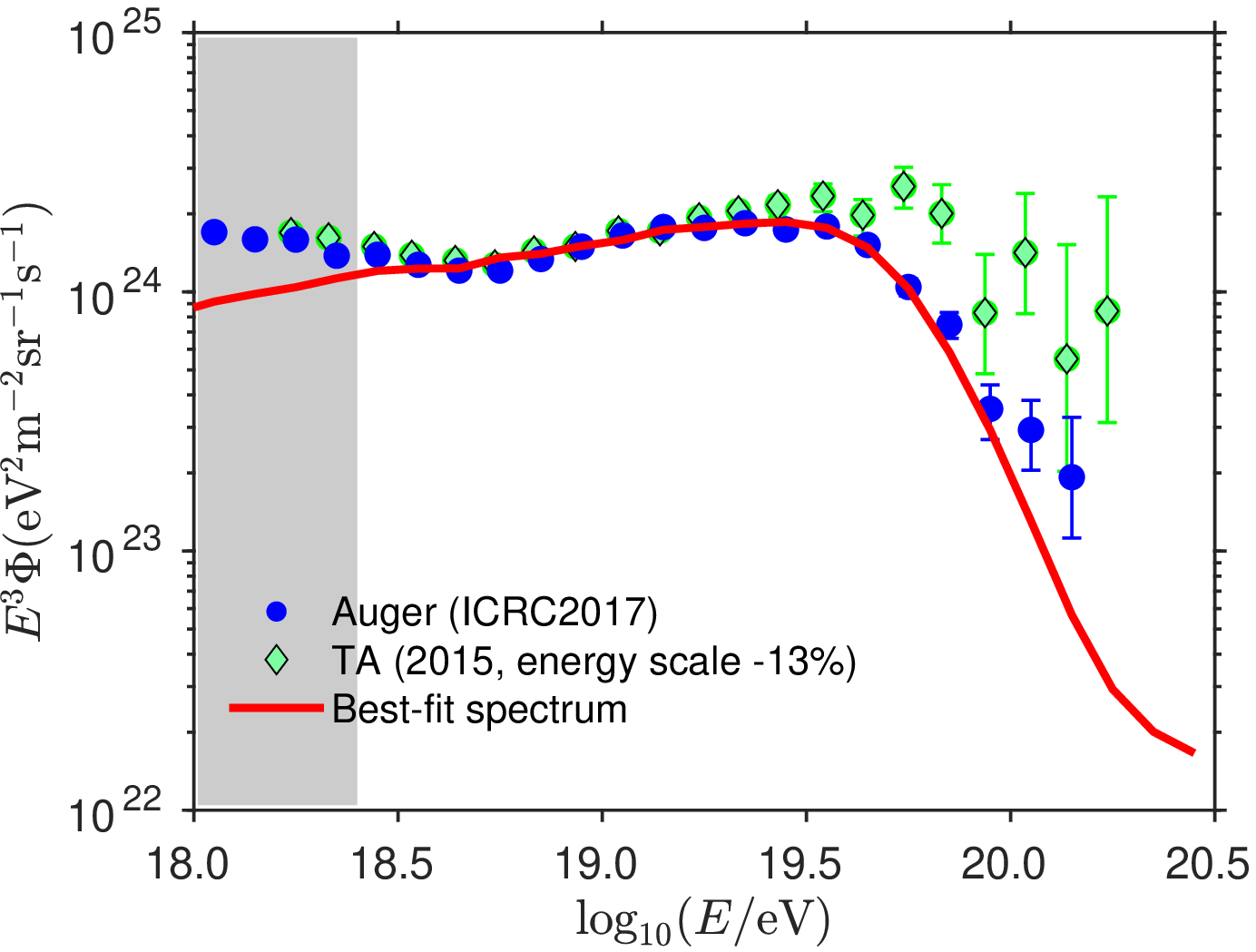} }}
\qquad
\subfloat[Spectral fit for Si]{{\includegraphics[width=0.5\linewidth]{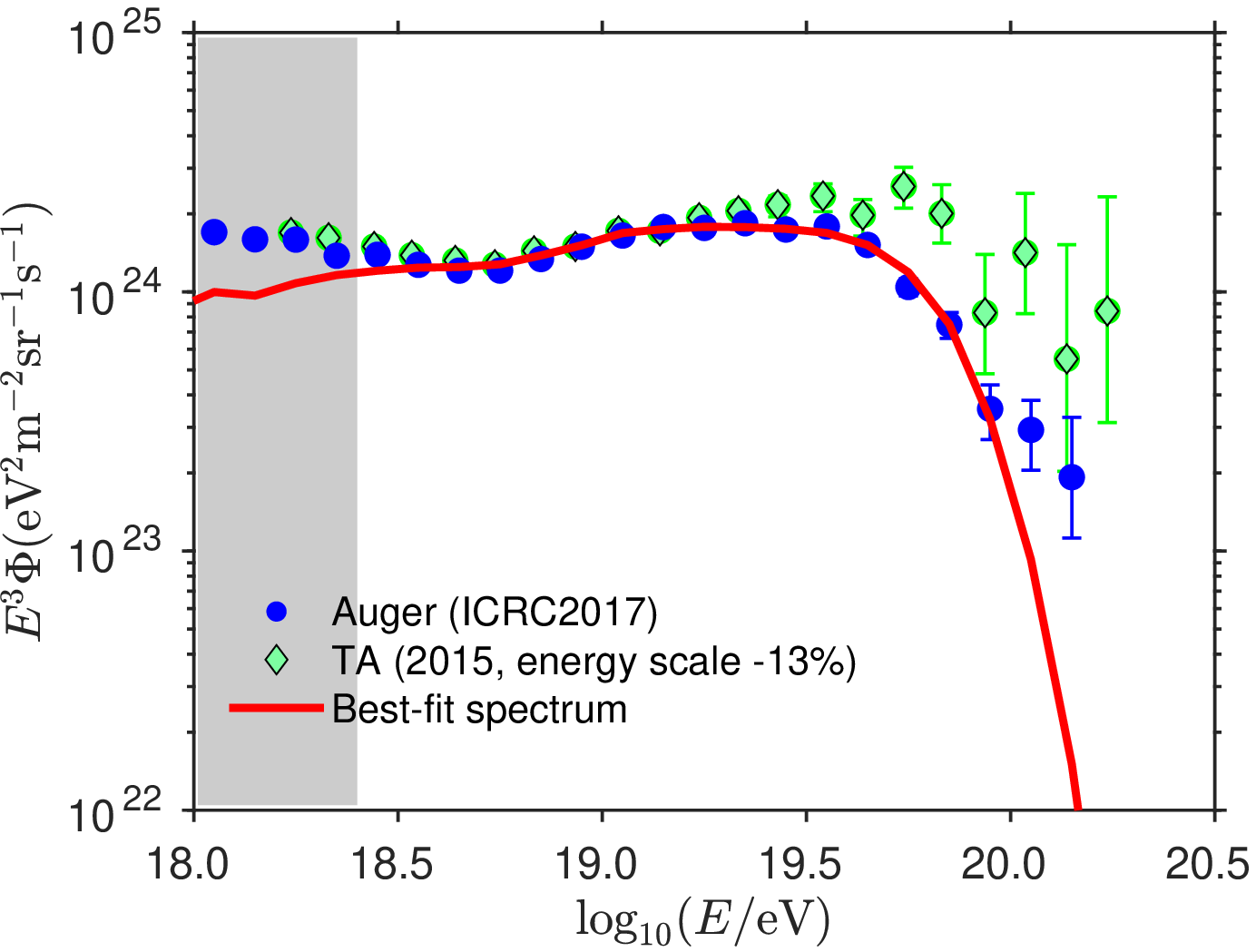} }}
\subfloat[Spectral fit for Fe]{{\includegraphics[width=0.5\linewidth]{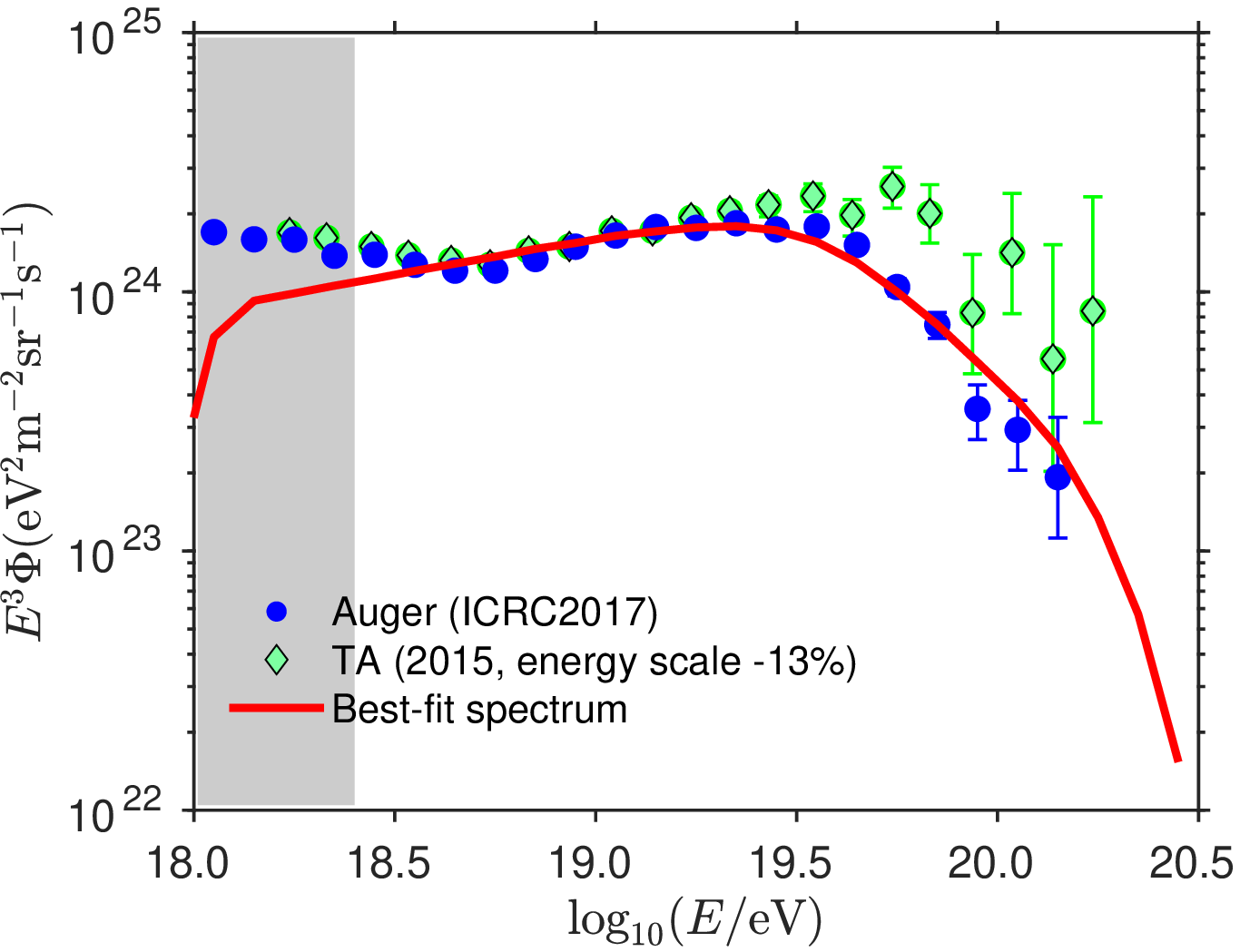} }}
\caption{Our best-fit spectra for pure He, O, Si and Fe injections, where only $\chi_{\rm spec}^2$ is considered. The best-fit parameters for He, O, Si and Fe are ($s=1.8, m=6, E_{\rm max}=10^{20.5}\rm eV$), ($s=1.6, m=7, E_{\rm max}=10^{20.9}\rm eV$), ($s=0.1, m=7, E_{\rm max}=10^{19.6}\rm eV$), and ($s=2.7, m=-2, E_{\rm max}=10^{20.5}\rm eV$).
}
\label{fig:NucleiSpecBestFit}
\end{figure*}

In this section, we fit the observed energy spectrum with light and intermediate nuclei.
For simplicity, we choose three typical nuclear species: He, O and Si.
The results are shown in Fig.~\ref{fig:NucleiSpecBestFit}, where the corresponding minimum chi-square values are $\chi_{\rm min}^2/\rm d.o.f =$ 0.94 (He), 1.45 (O) and 1.24 (Si), respectively.
We can see the pure light and intermediate nuclear composition can give a reasonable fit to the observed energy spectrum.

\begin{figure*}%
\centering
\subfloat[Fitting range from $10^{18.45}$~eV to $10^{20.15}$~eV for He injection]{{\includegraphics[width=0.5\linewidth]{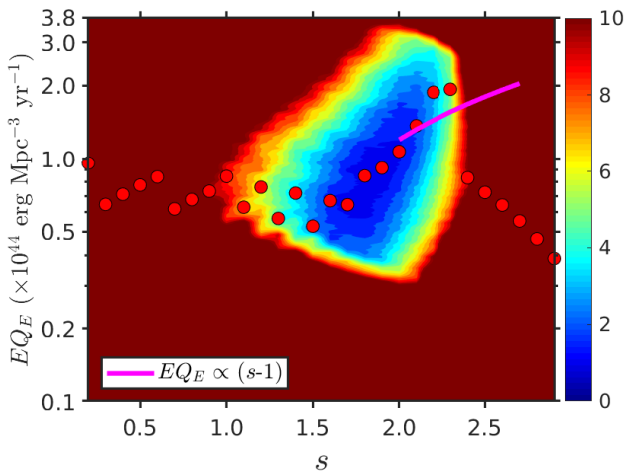} }}%
\subfloat[Fitting range from $10^{18.45}$~eV to $10^{20.15}$~eV for O injection]{{\includegraphics[width=0.5\linewidth]{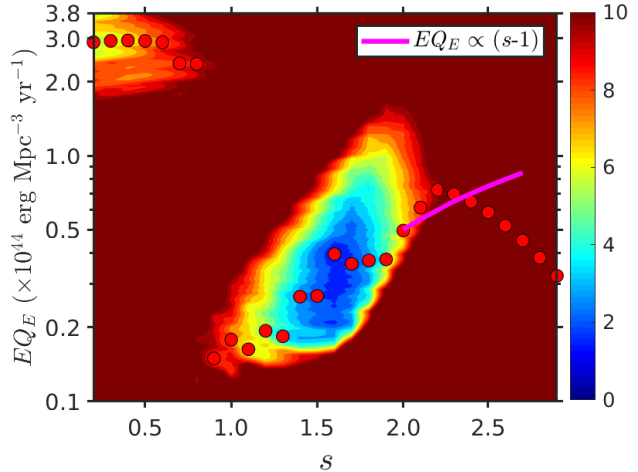} }}%
\qquad
\subfloat[Fitting range from $10^{18.45}$~eV to $10^{20.15}$~eV for Si injection]{{\includegraphics[width=0.5\linewidth]{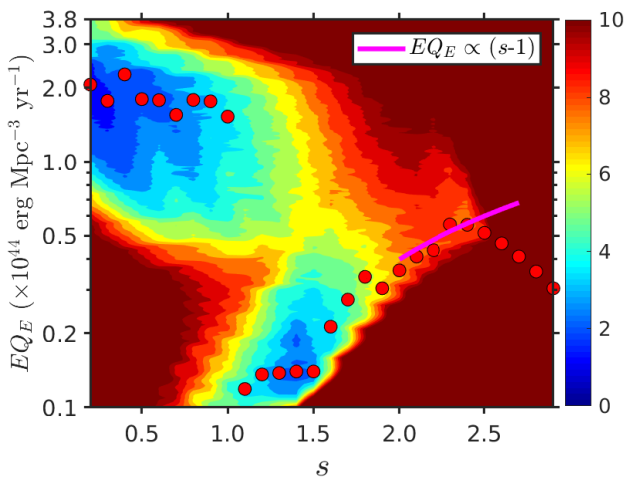} }}%
\subfloat[Fitting range from $10^{18.45}$~eV to $10^{20.15}$~eV for Fe injection]{{\includegraphics[width=0.5\linewidth]{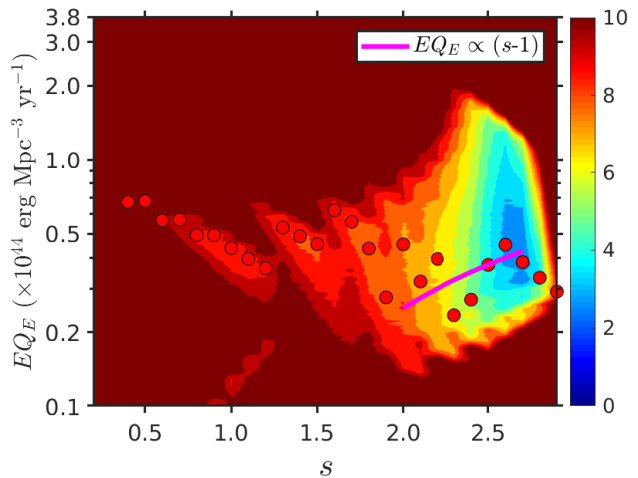} }}%
\caption{Best-fit UHECR generation rate density $EQ_E^{19.5}$ as a function of the spectral index $s$ (red solid dots), compared with the analytic dependence of $EQ_E^{19.5} \propto s-1$ in the $s$ range of [2.0, 2.7]. As in Fig.~\ref{fig:injectionWithGammaProton}, the background contours show $\chi_{\rm min}^2$/d.o.f., and different species of nuclei at the sources are assumed in different subplots.}%
\label{fig:injectionWithGammaNuclei}%
\end{figure*}

\begin{figure*}%
\centering
\subfloat[${(\chi_{\rm min}^2/{\rm d.o.f.})}_{\rm spec}$ for O injection]{{\includegraphics[width=0.33\linewidth]{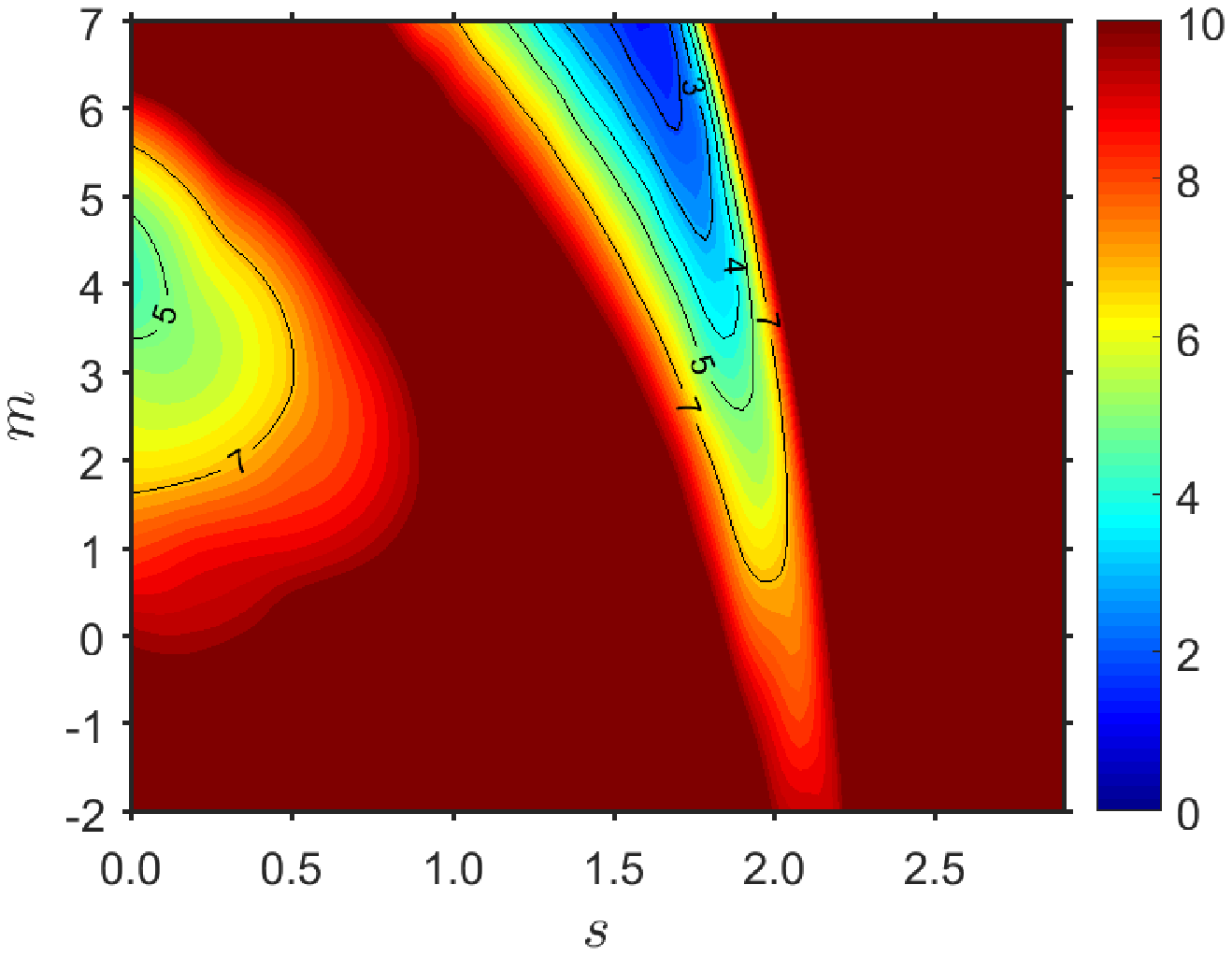} }}%
\subfloat[${(\chi_{\rm min}^2/{\rm d.o.f.})}_{\rm spec}$ for Si injection]{{\includegraphics[width=0.33\linewidth]{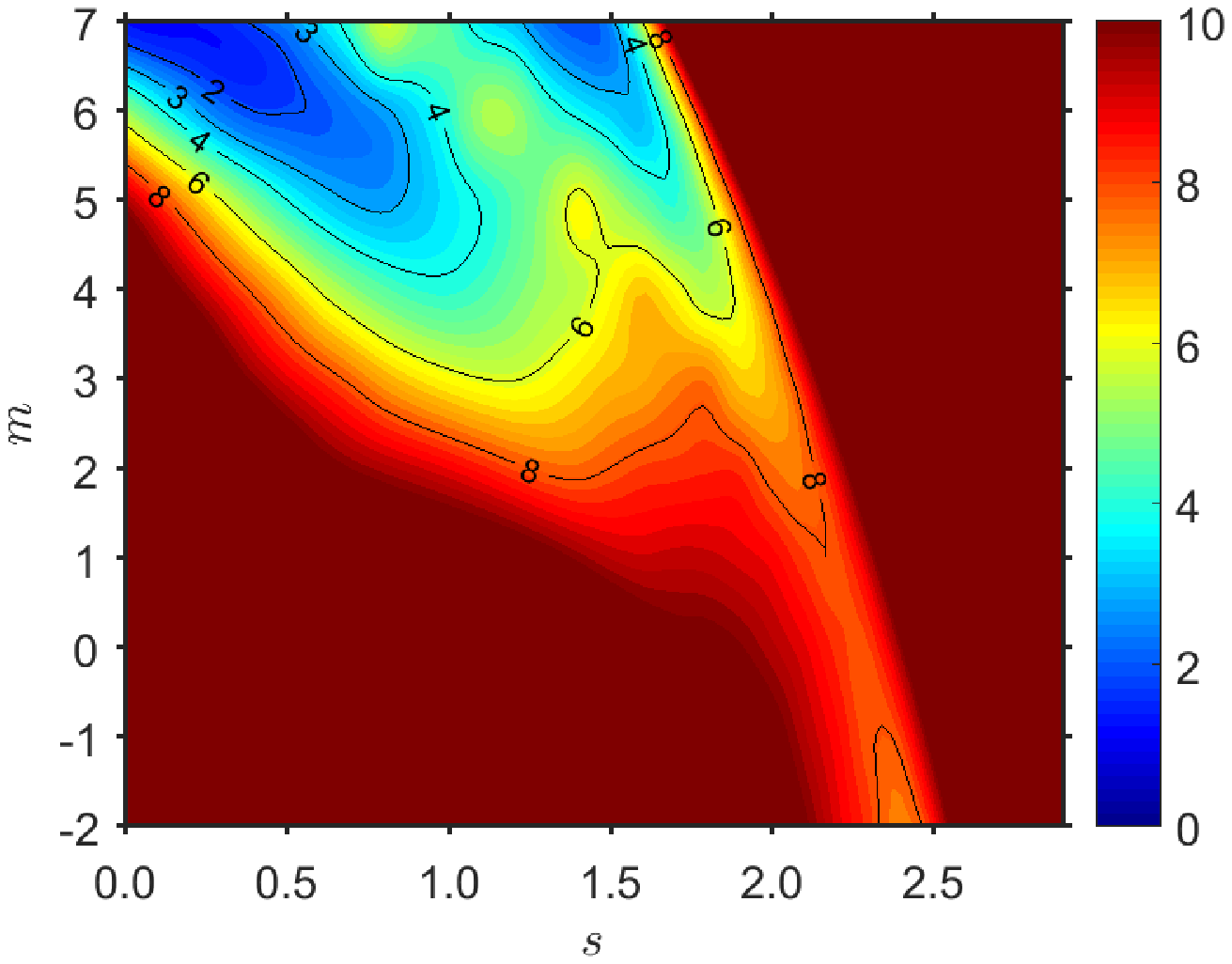} }}%
\subfloat[${(\chi_{\rm min}^2/{\rm d.o.f.})}_{\rm spec}$ for Fe injection]{{\includegraphics[width=0.33\linewidth]{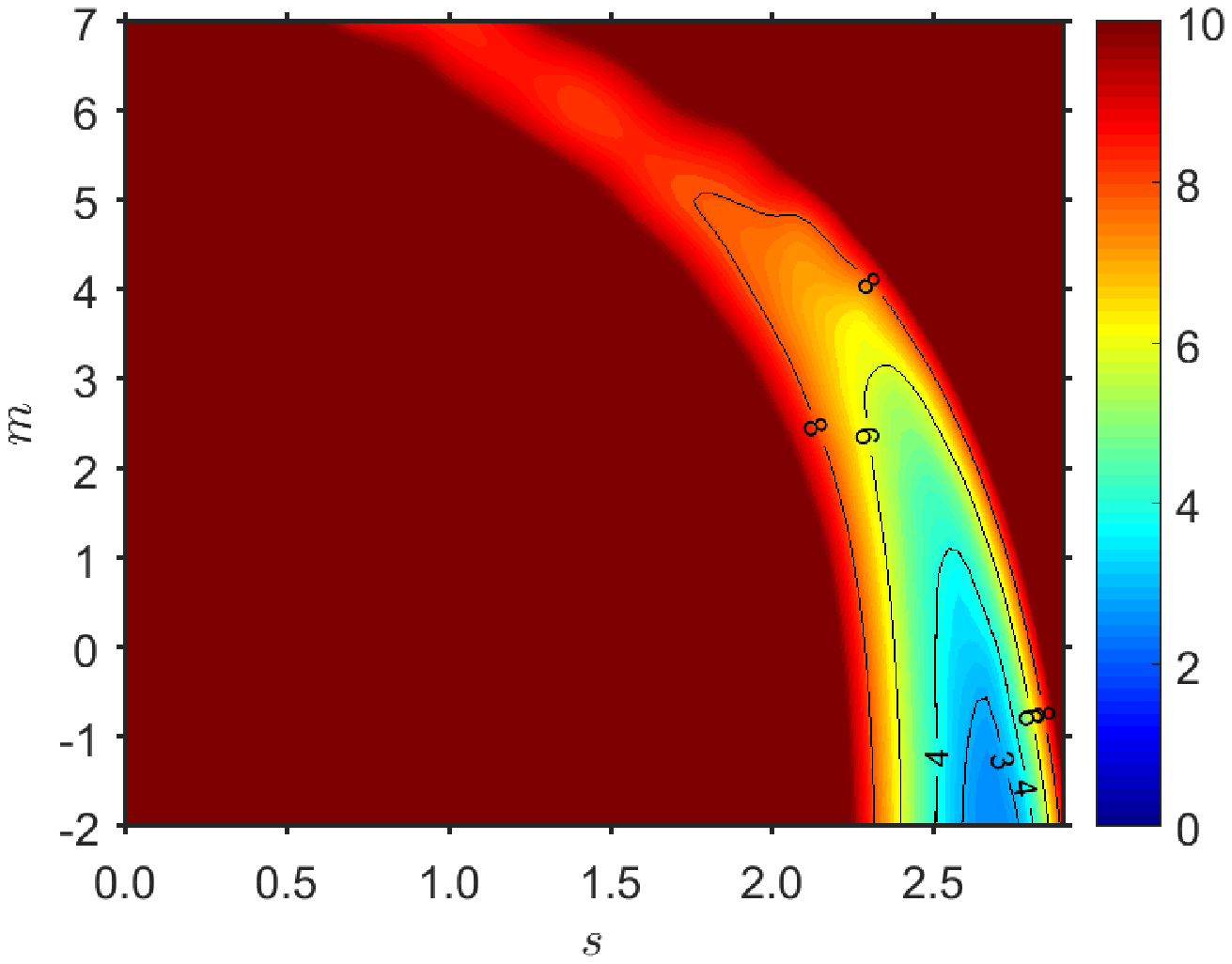} }}%
\caption{\add{As in Fig.~\ref{fig:3DscanProton}(b), the best-fit parameter space for the spectral index $s$ and redshift evolution index $m$ are shown but for the injection of nuclei rather than protons. The contours represent $\chi_{\rm min}^2$/d.o.f., where $E_\text{max}$ is optimized for the fitting.}}%
\label{fig:3DscanSiFe}%
\end{figure*}

The injection of light and intermediate nuclei generally does not yield spectral fits as good as proton case, especially for nuclei with larger mass $A$ which usually have larger values of $\chi^2/\rm d.o.f$. This indicates that the spectrum is composed of the mixture of protons and some heavy nuclei, or that proton and light nuclei are still dominant especially around the ankle energy as shown by some previous studies~\cite{Aab:2016zth}.
In the Si case of Fig.~\ref{fig:3DscanSiFe}, there are two local minima in the $\chi^2$ distribution as a function of $s$ and $m$. 
The one with the best-fit parameter set has a spectral power-law index ($s$) close to 0 and strong redshift evolution corresponding to a larger value of $m$. The region of lower values of $\chi^2$ containing this minima extends towards $s\sim1$ and $m\sim4$, and larger values of $s$ correspond to lower values of $m$, which is a general feature for different species of nuclei. 
The other with a less extended local minimum has a larger value of $\chi^2$ and its central value of $s$ is a little bit smaller than 1.5, corresponding to a softer spectrum than the best-fit one. However, as shown in Fig.~\ref{fig:3DscanSiFeWithXmax}, the fit to both the spectrum and $X_\text{max}$ is not good in this region.

\subsection{Heavy nuclei}
\begin{figure}
\includegraphics[width=1.0\linewidth]{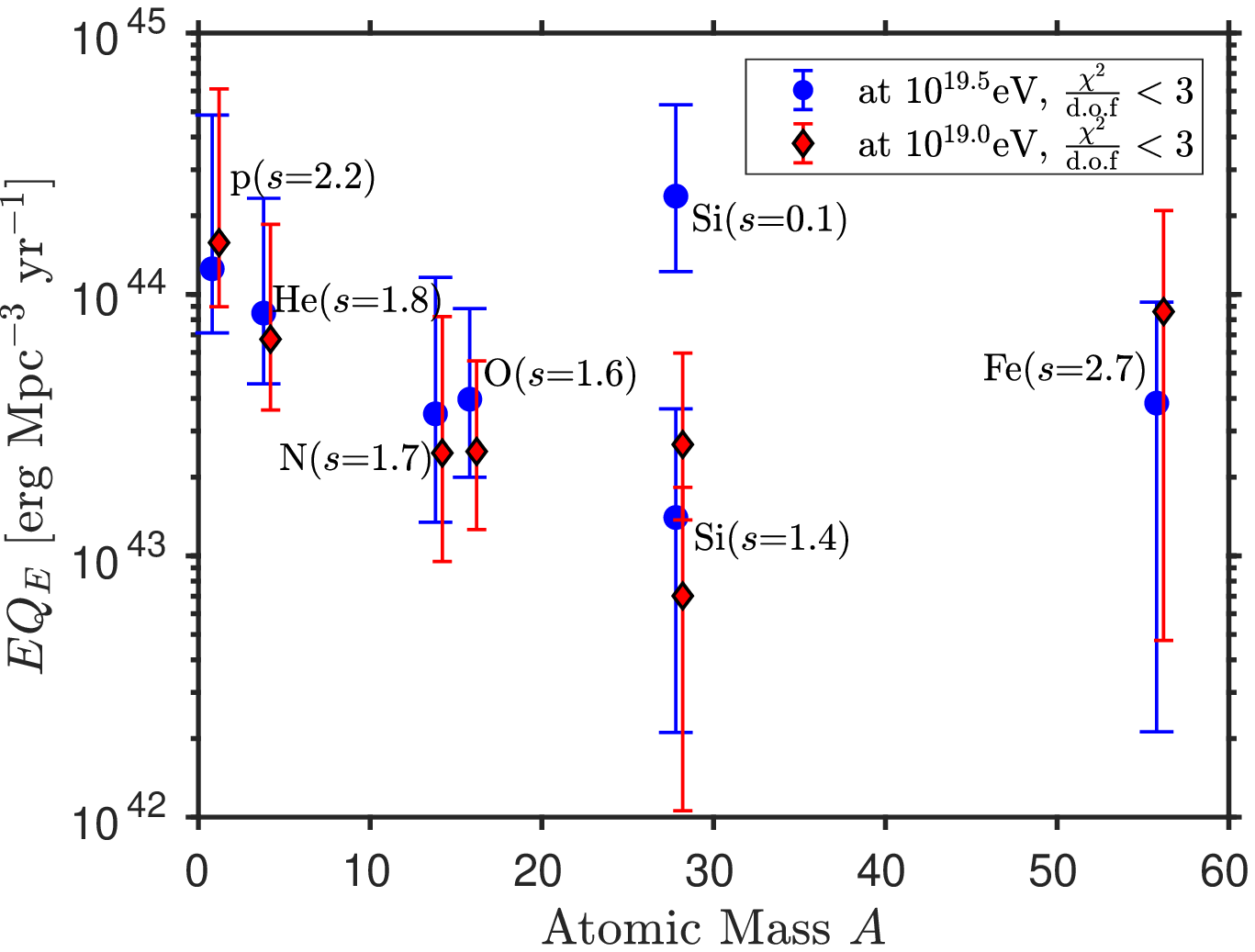}
\caption{
The differential UHECR energy generation rate density $EQ_E$ (at the reference energy of $10^{19.5}$~eV and $10^{19.0}$~eV) as a function of the mass number $A$, where we consider p, He, N, O, Si and Fe. Parameters $s$, $m$ and $E_\text{max}$ are scanned, and one or two values of $s$ that give the lowest $\chi^2$ are indicated. To show the ``conservative'' range of the energy generation rate density, values allowed by $\chi^2/\rm d.o.f.<3$ are depicted as error bars.
} 
\label{fig:injVSmass}
\end{figure}
We consider Fe as an example of the injection of heavy nuclei. The energy loss length of photodisintegration of Fe is around 100 Mpc for Fe with energy around $10^{20}$~eV, which is comparable to the energy loss length of proton and larger than those of intermediate nuclei around $10^{20}$~eV. So we choose Fe to be the heaviest nuclei we examine and present its results separately.
We obtain $\chi_{\rm min}^2/\rm d.o.f =2.57$ for Fe, so it does not have an excellent spectral fit and it is actually the worst among the nuclei we examine. Ref.~\cite{Aab:2016zth} presented the composition analysis that leads to the best fit, in which Fe fraction was zero. 
On the other hand, as shown in Fig.~\ref{fig:NucleiSpecBestFit}, Fe is the only pure composition to fit well the highest-energy spectral data points, which might indicate it plays a role if we fully examine the mixed composition scenario. It is worthwhile to mention $E_\text{max}\sim10^{20.5}$eV in the best spectral fit of Fe, which is not so high. This is also the case for lighter nuclei. Thus, it is likely that the intrinsic maximum energy at the sources and the cutoff energy due to the propagation both play important roles in shaping the observed UHECR spectrum. 
The best-fit value of $s=2.7$ for Fe indicates a soft injection spectrum. On the other hand, the best-fit value of $s=0.1$ for Si, which is a very hard spectrum and inconsistent with what we expect from first order Fermi shock acceleration. For all other species of nuclei we test, the best-fit values of $s$ are between 1.5 and 2.0. 

We note that some previous studies~\cite{Taylor:2015rla,Aab:2016zth} favored negative redshift evolution (i.e, smaller values of $m$). A part of the reason is that we use the pure composition of nuclei at the injection for our conservative purpose of studying the energy generation rate density. As shown in the next section, we also test the mixed composition used in Ref.~\cite{Aab:2016zth}, and we do end up with a similar $m$ preference. 

Last but not least, after testing above different species of nuclei from light to heavy, we are able to plot the energy generation rate density $EQ_E$ with $\chi_{\rm min}^2$ as a function of injected atomic mass in Fig.~\ref{fig:injVSmass}. The error bars represent the range of generation rate density that can lead to a spectral fit with $\chi^2/\rm d.o.f<3$, which corresponds to a p-value of 0.003\%. 
We also present $EQ_E$ at $10^{19.0}$~eV in the plot for comparison.
For the helium composition, we obtain a larger value of $EQ_E^{19.5}\approx0.85\times$10$^{44}$ erg Mpc$^{-3}$ yr$^{-1}$.
For oxygen and silicon that are intermediate nuclei, we get $EQ_E^{19.5}\approx0.40\times$10$^{44}$ erg Mpc$^{-3}$ yr$^{-1}$ and $EQ_E^{19.5}$ $\approx2.4\times$10$^{44}$ erg Mpc$^{-3}$ yr$^{-1}$, respectively. 
For the iron composition, we obtain $EQ_E^{19.5}\approx0.38\times$10$^{44}$ erg Mpc$^{-3}$ yr$^{-1}$. 
For heavy nuclei, the UHECR energy budget is smaller than that for the proton composition except for the best-fit case with silicon (that has a very hard spectrum with $s\sim0.1$).
Overall, for nuclei, we obtain $EQ_E^{19.5}\approx(0.4-1.5)\times$10$^{44}$ erg Mpc$^{-3}$ yr$^{-1}$ for $s\gtrsim1.5$, which is similar to the values of the proton case~\cite{Waxman:1995dg,Bahcall:2002wi,2008AdSpR..41.2071B,Katz:2008xx}. Although the value of $EQ_E^{19.5}$ in the cases of nuclei can be somewhat smaller if $s\lesssim1.5$, \km{we conclude that the UHECR energy generation rate density is insensitive to details of the composition within moderate uncertainties, and our results give insight into UHECR energetics independent of the composition measurements through the distribution of $X_\text{max}$.}

\section{Results of spectral and composition fits}
\begin{figure*}
\centering
\subfloat[Combined best fit ($\chi^2$ considering both the spectral and $X_\text{max}$ data) for O]{{\includegraphics[width=0.5\linewidth]{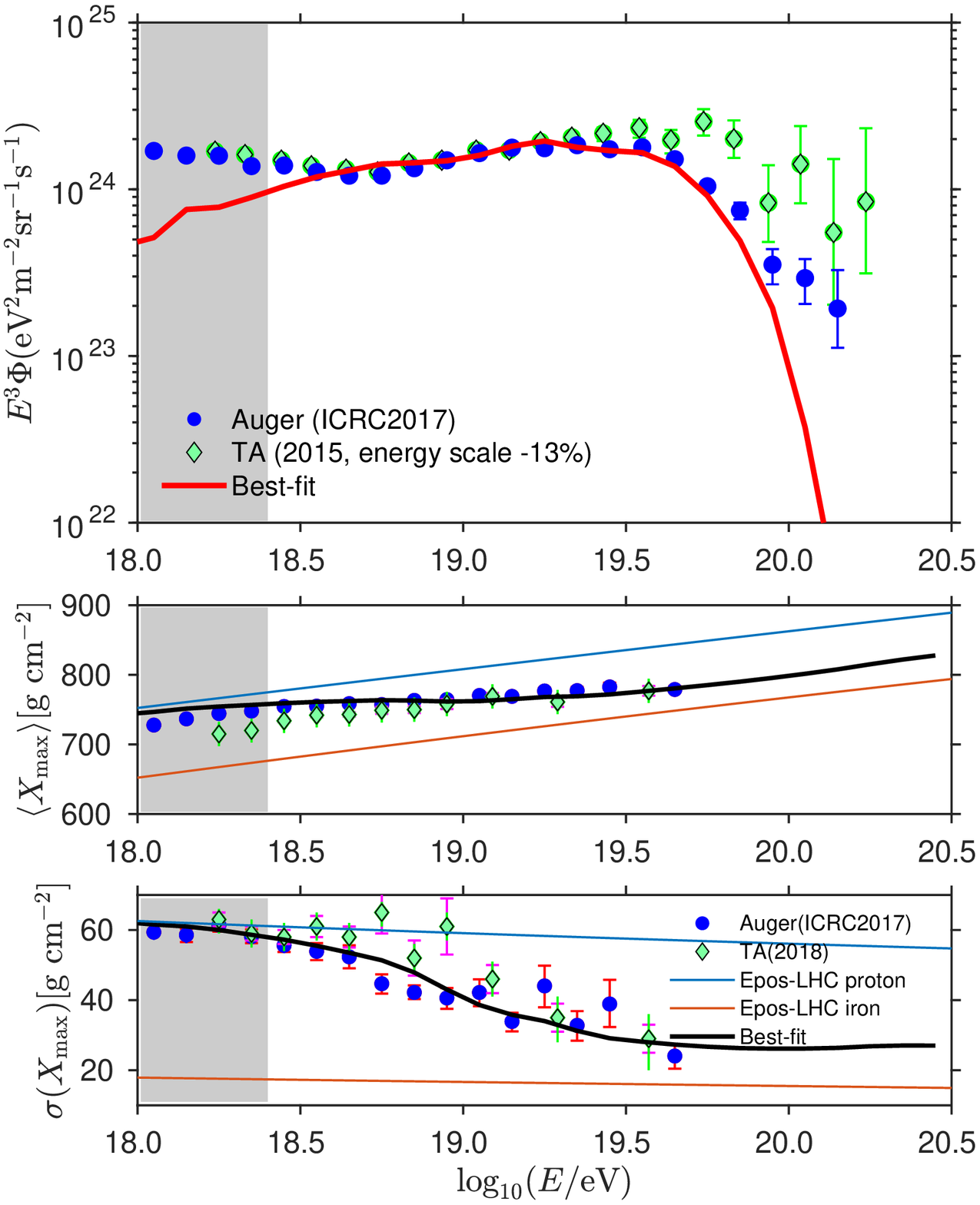} }}
\subfloat[Combined best fit ($\chi^2$ considering both the spectral and $X_\text{max}$ data) for Si]{{\includegraphics[width=0.5\linewidth]{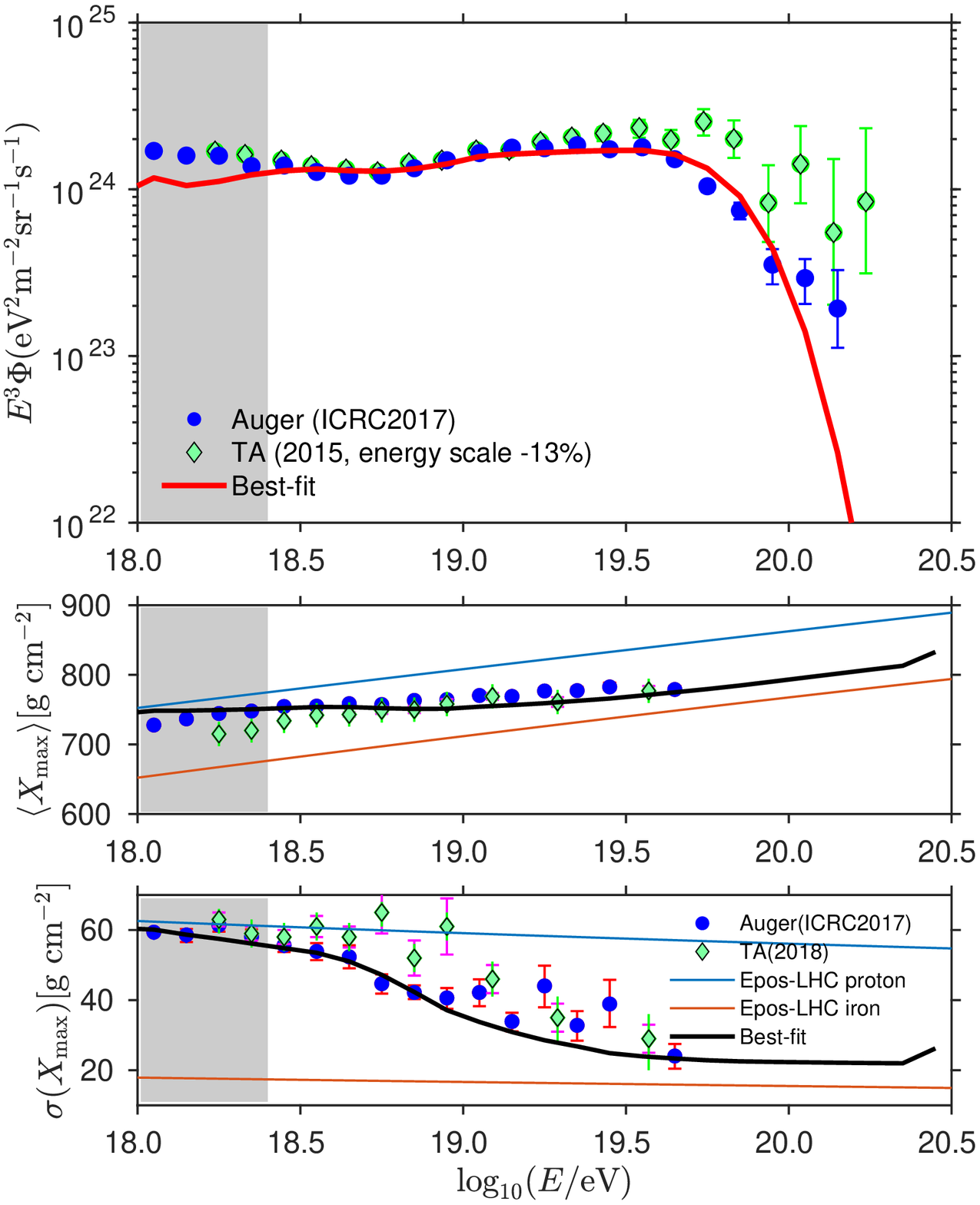} }}
\caption{Combined fits of pure O and Si simulated injection. Both the spectral and $X_\text{max}$ data are included in the search of best fits. In the average and standard deviation of the $X_\text{max}$ distribution subplots, the calculated $X_\text{max}$ moments from simulated spectra are compared to the experimental data. Predicted results from pure proton and iron are also showed for comparison. The hadronic interaction model in the air shower is assumed to be EPOS-LHC.
}
\label{fig:SiOXmaxFit}
\end{figure*}

\begin{figure*}%
\centering
\subfloat[${(\chi_{\rm min}^2/{\rm d.o.f.})}_{\rm total}$ for O injection]{{\includegraphics[width=0.33\linewidth]{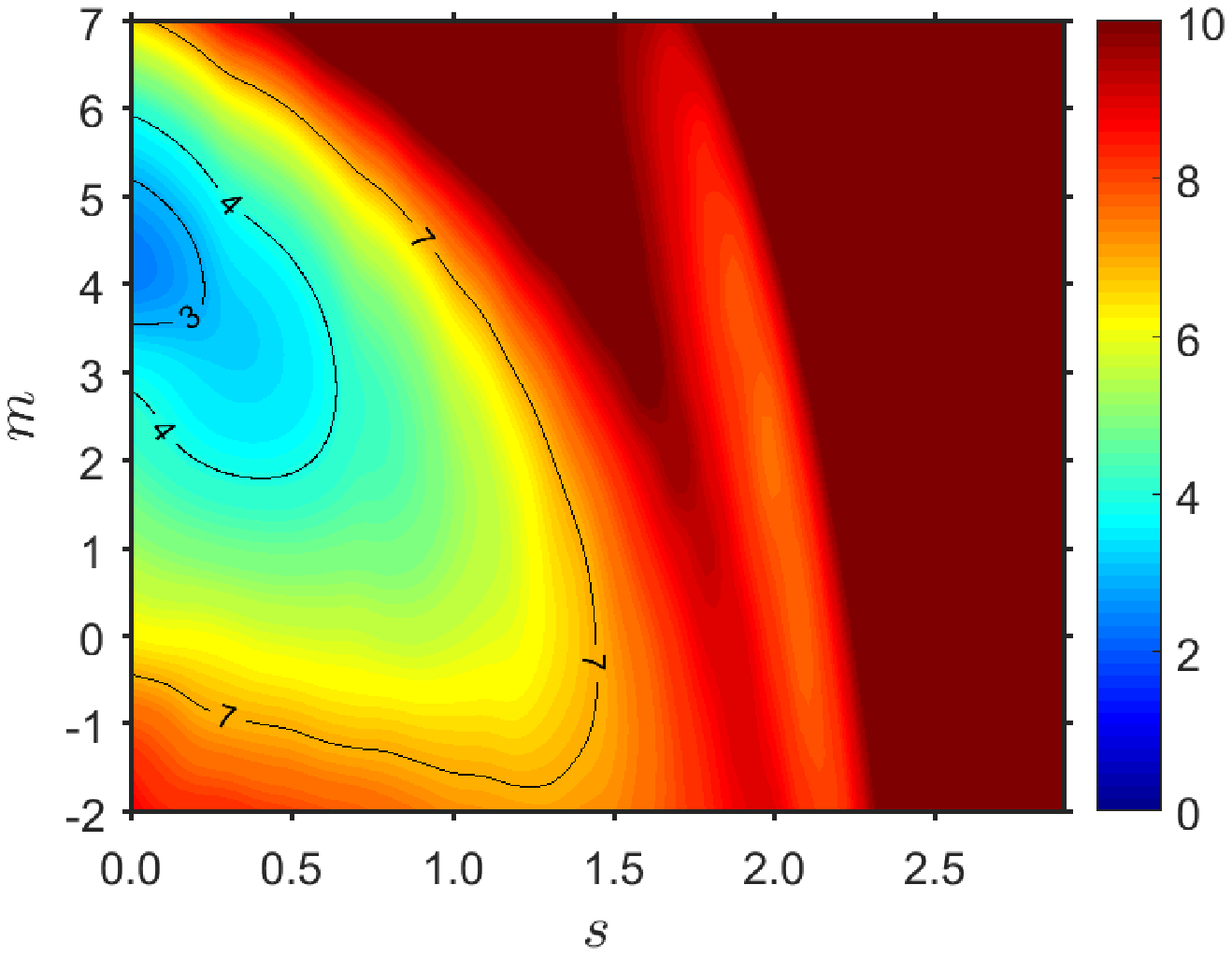} }}
\subfloat[${(\chi_{\rm min}^2/{\rm d.o.f.})}_{\rm total}$ for Si injection]{{\includegraphics[width=0.33\linewidth]{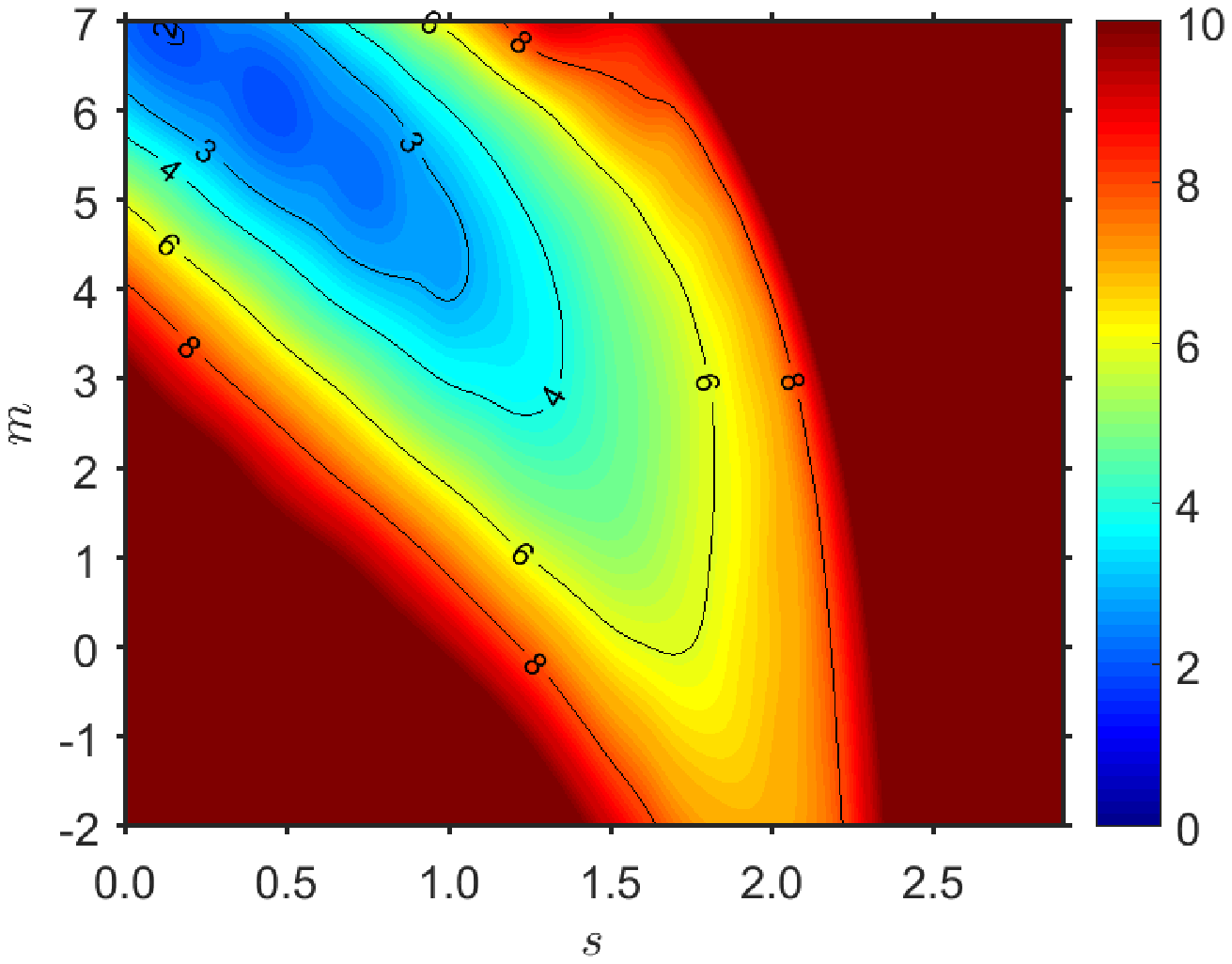} }}
\subfloat[${(\chi_{\rm min}^2/{\rm d.o.f.})}_{\rm total}$ for Fe injection]{{\includegraphics[width=0.33\linewidth]{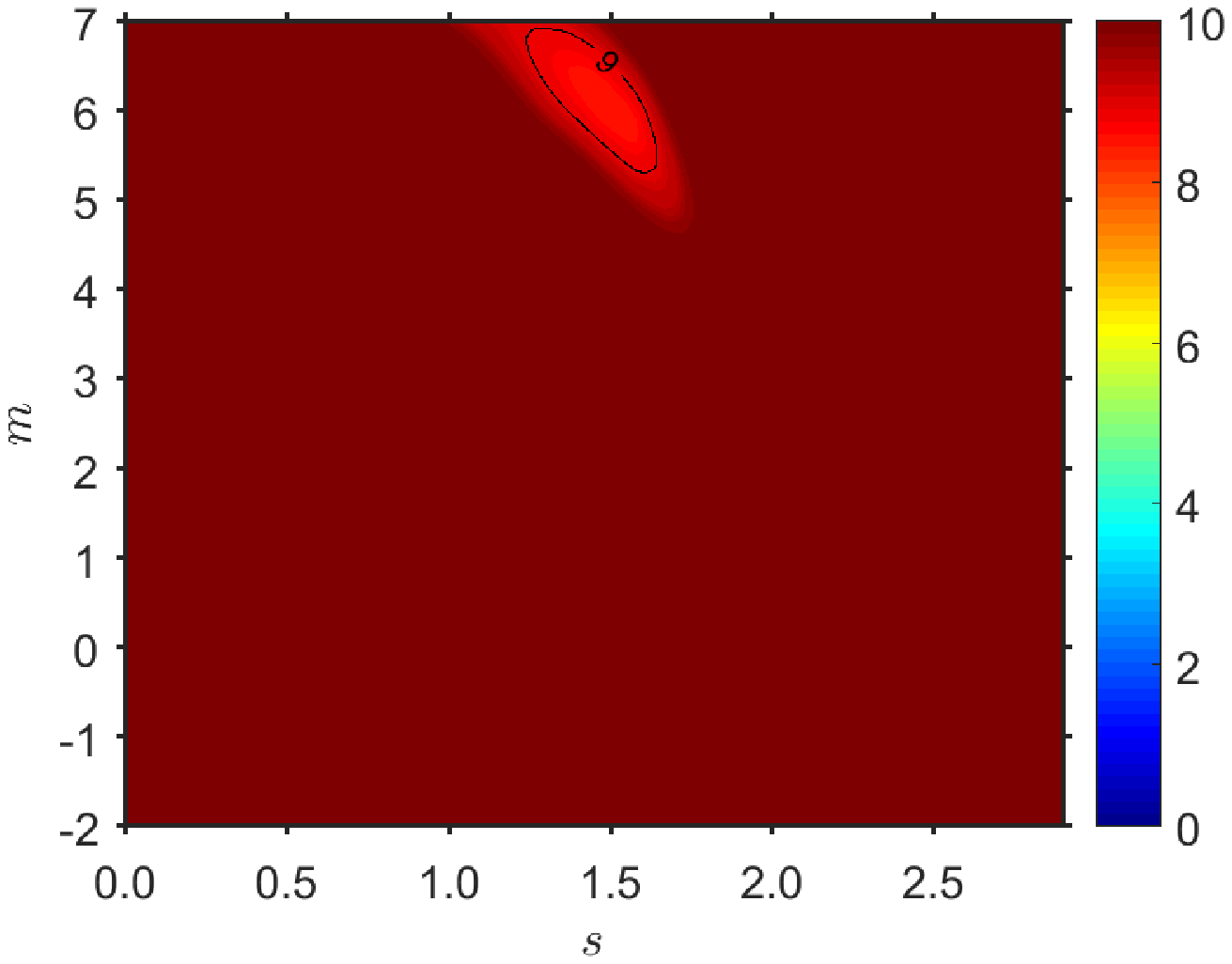} }}
\caption{
As in Fig.~\ref{fig:3DscanProton}, the best-fit parameter space for the spectral index $s$ and redshift evolution index $m$ is shown but for the injection of nuclei. The contours represent {($\chi_{\rm min}^2$/d.o.f.)}$_{\rm total}$, where $E_\text{max}$ is scanned.
The left subplot shows the result for pure O injection, and the middle subplot shows the result for pure Si injection, while the right one does for pure Fe injection.}%
\label{fig:3DscanSiFeWithXmax}%
\end{figure*}

In this section, we take into account the effect of fitting other two observables, $\langle X_\text{max} \rangle$ and $\sigma(X_\text{max})$. Obviously, the combined fit of both the spectrum and $X_\text{max}$ can provide more information about the composition of UHECR and help justify its origin and propagation models. The hadronic interaction model involved in air shower simulations is set to be EPOS-LHC~\cite{Pierog:2013ria} for simplicity. 
\km{The other hadronic interaction models give similar energy dependence of $\langle X_\text{max} \rangle$ and $\sigma(X_\text{max})$. Uncertainties among the hadronic interaction models in fitting the UHECR data is discussed in several works (e.g., Refs.~\cite{Aab:2016zth,Romero-Wolf:2017xqe}). 
}
The fitting energy range for $\langle X_\text{max} \rangle$ and $\sigma(X_\text{max})$ is from $10^{18.45}$eV to $10^{19.65}$eV. \km{As in the results of spectral fits, the composition data from TA (2018) are shown~\cite{Abbasi:2018nun} for comparison to the Auger data. The TA suffer from limited statistics at the highest energies (especially above $10^{19}$eV), but we opt to present them as well because the interpretation of the $X_\text{max}$ data is still under debate.}

For Si, the parameter sets for their spectral best fits give $\langle X_\text{max} \rangle$ and $\sigma(X_\text{max})$ data decent fits as well, which is indicated in Fig.~\ref{fig:SiOXmaxFit}
That means adding fits of $\langle X_\text{max} \rangle$ and $\sigma(X_\text{max})$ will not change the best-fit parameters much compared to only fitting spectrum. 
Some regions in the parameter space have small values of total $\chi_{\rm min}^2$ (thus good fits) combining the spectral and $X_\text{max}$ fits as shown in Fig.~\ref{fig:3DscanSiFeWithXmax}, which is also the case for O although its spectral best-fit parameters are not close to combined best-fit parameters. However, for some other nuclei like Fe, the best-fit parameter set for the UHECR spectrum gives a poor fit to $\langle X_\text{max} \rangle$ and $\sigma(X_\text{max})$ data, implying that the combined fit cannot give a small value of total $\chi^2$, as the Fe case shown in Fig.~\ref{fig:3DscanSiFeWithXmax}.
Among different injected nuclei we examine, only Si can produce a very decent fit of both the spectrum and $X_\text{max}$, as long as we consider pure composition models. This is consistent with results of Ref.~\cite{Aab:2016zth}.

\add{In the Fe case, as shown in Figs.~\ref{fig:3DscanSiFe} and \ref{fig:3DscanSiFeWithXmax}, the 3D scan of parameters implies that the area with small values of spectral $\chi^2/\rm d.o.f$ is localized near the boundaries of parameters that we scan. Parameters giving small values of $\chi^2$ barely overlap between spectral and $X_\text{max}$ fits. }
That is the case for most of the species of injected nuclei that we consider. Another interesting feature in the 3D scan of parameters is that the combined fit essentially rules out $m < 3$ because these regions lead to large values of $\chi^2$, regardless of the composition at injection. Pure composition models would not work in many source classes with redshift evolution stronger than $m\sim3$. 

We compare our combined fits of spectrum and $X_\text{max}$ with the results of Ref.~\cite{Aab:2016zth}. The best-fit composition at the sources indicated in Ref.~\cite{Aab:2016zth} is dominated by He, N and Si. We adopt the same mixed composition of nuclei, where the mass fraction of He is 67.3\%, N is 28.1\% and Si is 4.6\%. We then apply the best-fit (including the combined fit of spectrum and $X_\text{max}$) parameters used by the Auger Collaboration, $s=0.96$, $E_\text{max}/Z=10^{18.68}$~eV with $m=0$, and we confirm that these lead to well-fitted results using our simulation and analysis method.
We compare the integrated energy generation rate density, $Q = \sum_{A}\int_{E_{\text{min}}}^{+\infty} Eq_{A}(E)\text{d}E$, in which $E_{\rm min}=10^{15}$~eV, for the best-fit parameters. Here $q_{A}(E)$ means the nuclei quantity of mass $A$ injected in unit energy, volume and time~\cite{Aab:2016zth}.
We obtain $Q\simeq5.3\times10^{44}~\text{erg Mpc}^{-3} \text{yr}^{-1}$, which is consistent with $4.99\times10^{44}~\text{erg Mpc}^{-3} \text{yr}^{-1}$ obtained by Ref.~\cite{Aab:2016zth}. 
Each contribution from different species of nuclei is similar as well. Note that Ref.~\cite{Aab:2020rhr} also gives the similar luminosity density $\simeq6\times10^{44}\text{erg Mpc}^{-3} \text{yr}^{-1}$ above $5\times10^{18}~$eV. 
For further comparison, we also scan the parameters to get $Q$ for our best-fit parameters. We find that values of $Q$ obtained with our simulations are $\gtrsim50$\% larger than the result with the best-ft composition in Ref.~\cite{Aab:2016zth}, both of the results are consistent within uncertainties (within a factor of 2). Some discrepancy is possible, because we fit the mean and RMS of $X_\text{max}$ data while Ref.~\cite{Aab:2016zth} fitted the whole distribution of $X_\text{max}$. 
Nevertheless, the calculated energy generation rate densities are similar between our results and Auger results. Furthermore, they are consistent with those from the spectral fits.


\section{Summary and discussion}
We evaluated the UHECR energy generation rate density for different species of nuclei injected at the sources. 
We used the public tool \textsc{CRPropa 3} to simulate the propagation of UHECR protons and nuclei, and fit the results with the UHECR spectrum measured by Auger. We also utilized the mass composition data as additional information.

We scanned three parameters characterizing the sources, the spectral index of UHECRs escaping from the sources ($s$), the maximal energy at the sources ($E_\text{max}$), and the index of redshift evolution ($m$). Besides the systematic uncertainty in the UHECR data, there are various uncertainties that can impact the results, including the fitting energy range, hadronic interaction models in EAS simulations, and propagation models of UHECRs. Nevertheless, we obtained some general features and correlations among the parameters ($s$, $E_\text{max}$, $m$, and $EQ_E^{19.5}$).

For the proton composition, the best-fit parameters indicate that an almost flatter energy spectrum, stronger redshift evolution and lower maximal energy at the sources give a better explanation for the spectrum measured by Auger. This result is also consistent with the previous analysis by Ref.~\cite{Aab:2016zth}. For the pure proton injection, the spectral fit itself is overall good (with lower values of $\chi_{\rm min}^2$).
The resulting UHECR energy generation rate density is $EQ_E^{19.5}\approx(0.7-2.0) \times$10$^{44}$ erg Mpc$^{-3}$ yr$^{-1}$ depending on the fitting energy range (see Fig.~\ref{fig:injectionWithGammaProton} and Fig.~\ref{fig:Proton3DInjection}), which is consistent with previous works~\cite{Murase:2008sa,Katz:2008xx,Waxman:1995dg,Bahcall:2002wi,2008AdSpR..41.2071B}. We also considered different species of heavier nuclei at the sources, and evaluated the energy generation rate densities via spectral fits (see Fig.~\ref{fig:injVSmass} for a summary), which also agrees with $EQ_E^{19.5}\approx(0.3-2.0)\times$10$^{44}$~erg~Mpc$^{-3}$ yr$^{-1}$ in Ref.~\cite{Murase:2018utn}. 

We found that with the best-fit parameters of the combined fits to the spectrum and composition, the goodness of fit is poorer for pure composition. \add{This indicates mixed composition models at the sources would be more reasonable, which is in agreement with several  analyses~\cite{Aab:2016zth,Aloisio:2013hya,Taylor:2013gga,Zhang:2017hom,Kimura:2017ubz,Zhang:2017moz,Zhang:2018agl}.} The pure proton injection can give a good fit only for the spectrum but not for the combined data, and with known hadronic interaction models the Auger composition data imply that the composition of UHECRs is heavier than protons~\cite{ThePierreAuger:2015rha}.

Although we do not focus on the detailed discussion on such a mixed composition interpretation, we stress that our results relying on the spectral fits are already conservative for the purpose of evaluating $EQ_E$, because $X_{\rm max}$ gives only additional constraints. 
We obtained that $EQ_E^{19.5}\approx(0.4 - 2.0)\times$10$^{44}$~erg~Mpc$^{-3}$~yr$^{-1}$, for arbitrary composition models. 
For a given hadronic interaction model, the combined fit of the spectrum and composition requires a mixed composition or pure Si-like injection, in which the energetics requirement lies within the range of our results. 
\km{Our results on energetics can be compared} with the energetics of different sources, such as gamma-ray bursts (GRBs), tidal disriuption events (TDEs), active galactic nuclei (AGN), galaxy clusters and so on (see TABLE II of~\cite{Murase:2018utn} for a summary), to narrow down the possible origins of UHECRs. The values we obtained are comparable to luminosity densities of extreme astrophysical sources. 
\km{The energetics constraint can be satisfied by many source classes, but viable scenarios are limited when the Hillas condition is taken into account. GRBs and energetic supernovae (including hypernovae and low-luminosity GRBs), jetted TDEs, AGN with jets (blazars and radio galaxies), starburst galaxies, and galaxy clusters are the most promising. 
Our results can be regarded as the most conservative estimates on the UHECR generation rate density, which can also relax the energetics requirement for GRBs and jetted TDEs. On the other hand, if the composition is dominated by nuclei, our results imply that a large amount of nuclei have to be loaded in the sources whatever the spectral index is, which is theoretically challenging. One possibility is the massive stellar origin of heavy nuclei, and low-luminosity GRBs and engine-driven supernovae provide a natural solution~\cite{Zhang:2017moz,Zhang:2018agl}. Another possibility is reacceleration of galactic cosmic rays by jets or outflows~\cite{Kimura:2017ubz,Mbarek:2019glq}.}

In this work, we did not consider impacts of IGMFs, although they can in principle influence the results in different ways (e.g., Refs.~\cite{Sigl:2003ay,Aloisio:2004jda,Aharonian:2010va,Durrer:2013pga,Das:2008vb}). 
\km{First, they allow the injection spectral index $s$ to be larger because the lower-energy particles will propagate longer in the magnetic fields and thus may lose more energies than the case without IGMFs~\cite{Lemoine:2004uw,Kotera:2007ca,Mollerach:2013dza,Wittkowski:2017okb,Fang:2017zjf}.} 
Second, our ``injection spectra, i.e., spectra of cosmic rays injected into intergalactic space'' should be regarded as ``spectra of cosmic rays leaving magnetized environments surrounding UHECR accelerators'' strictly speaking (see Ref.~\cite{Fang:2017zjf} as an example). 
However, in reality, the confinement in the magnetized environments (such as radio lobes and galaxies) can also make the difference between the spectrum of UHECRs leaving accelerators and that of UHECRs leaving the surrounding environments. 
Further three-dimensional simulations of propagation processes can be conducted if we consider the defection by IGMFs, which will make the received spectrum softer out of the same source set-up.
Third, the simple power-law assumption we take is generally applied to accelerated spectrum thus the injection spectrum can be further explored if we consider the confinement and other processes before the UHECRs escape. In such a set-up, the pure composition assumption of escaping UHECRs is not realistic~\cite{Unger:2015laa,Fang:2017zjf} and we must consider mixed composition models for spectra injected into intergalactic space, although the UHECR energy generation rate density should still fall into the range we give. 

Future observations will quantitatively improve our results on the energy generation rate density. One uncertainty in our fits of $X_{\text{max}}$ data is the choice of hadronic interaction models, and more sophisticated experiments may help us understand the physics of air showers more deeply.
With newly designed and developed experiments, such as the next-generation AugerPrime experiment~\cite{Aab:2016vlz}, the Probe Of Extreme Multi-Messenger Astrophysics (POEMMA)~\cite{Olinto:2017xbi} mission, and the Giant Radio Array for Neutrino Detection (GRAND)~\cite{Alvarez-Muniz:2018bhp}, the systematic error will then be reduced in an more extended energy scale. Together with other observables such as anisotropy in arrival directions of UHECRs, this will help constrain source models and have implications for the origins of UHECRs.


\medskip
\begin{acknowledgments}
The results of this work were first presented at ICRC 2019. 
While this work is being prepared, the new Auger data appeared~\cite{Aab:2020gxe}. However, our results on the UHECR energy generation rate density are essentially unaffected by the new data. 
This work is supported by the Alfred P. Sloan Foundation, NSF Grant No.~AST-1908689, and KAKENHI No.~20H01901 and No.~20H05852 (K.M.). B.T.Z. acknowledges the IGC fellowship. 
We thank Foteini Oikonomou for helpful comments on this manuscript. We also thank Glennys Farrar for useful comments at ICRC 2019. 
Computations for this research were performed on the Pennsylvania State University’s Institute for Computational and Data Sciences’ Roar supercomputer.
\end{acknowledgments}


\bibliography{ms}

\newpage
\appendix
\section{Impacts of the Fitting Range}\label{appendix:fit}
This section is devoted to the discussion on the impact of the different fitting energy range in the search of best-fit parameters. We here present the results of another fitting range from $10^{19.05}$~eV to $10^{20.15}$~eV for the purpose of comparison. 
It demonstrates that our results are robust considering the similarities between the results of different fitting ranges. For energies greater than $10^{19.05}$~eV, there may be spectral discrepancies between the Auger and TA data, so the results would be more robust for fitting range starts at $10^{18.45}$~eV.

\subsection{Spectral fits}

\begin{figure*}[tb]%
\centering
\subfloat[The best-fit spectrum for proton]{{\includegraphics[width=0.33\linewidth]{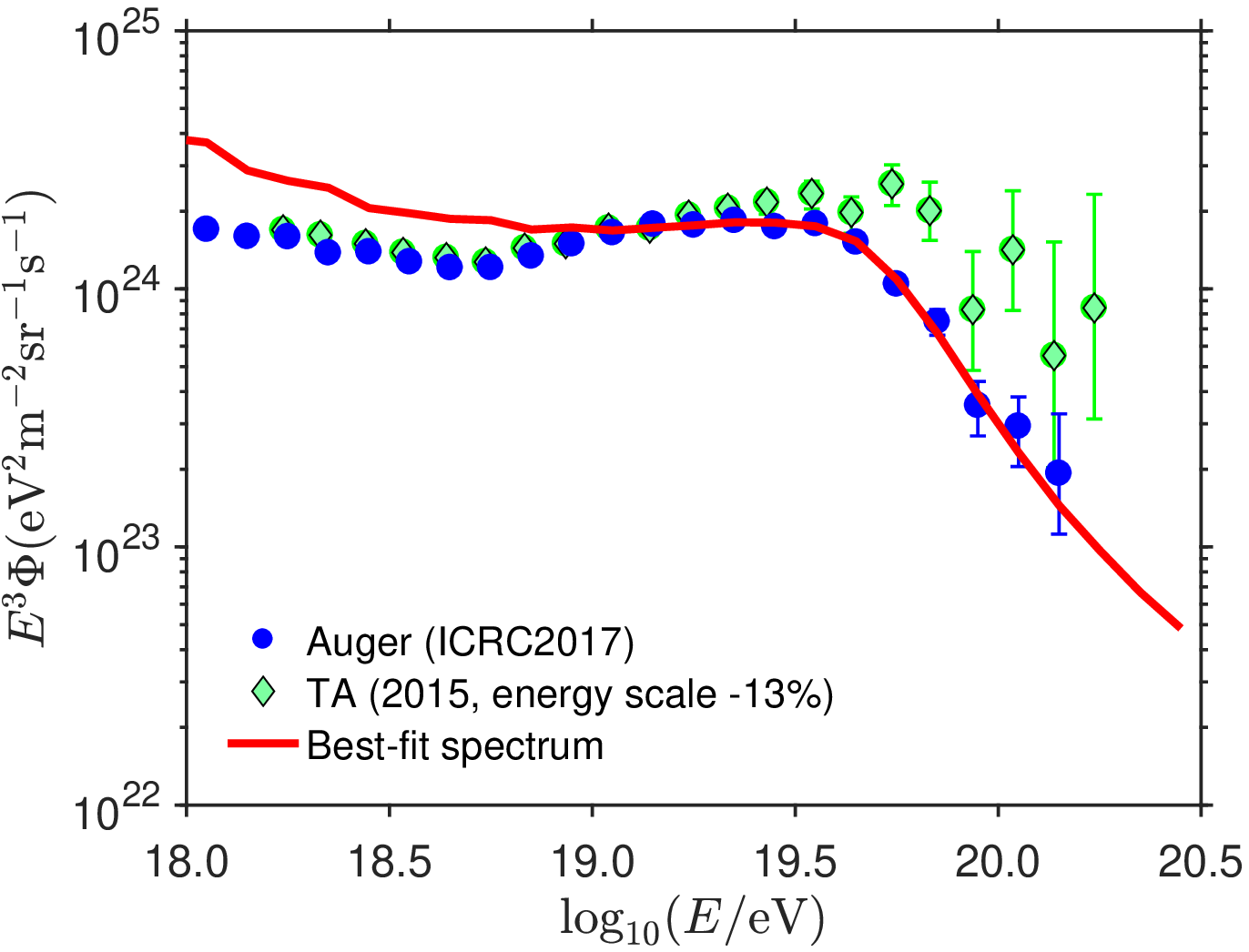}}}%
\subfloat[$\chi_{\rm min}^2$/d.o.f. as a function of $s$ and $m$, where $E_\text{max}$ is scanned.]{{\includegraphics[width=0.33\linewidth]{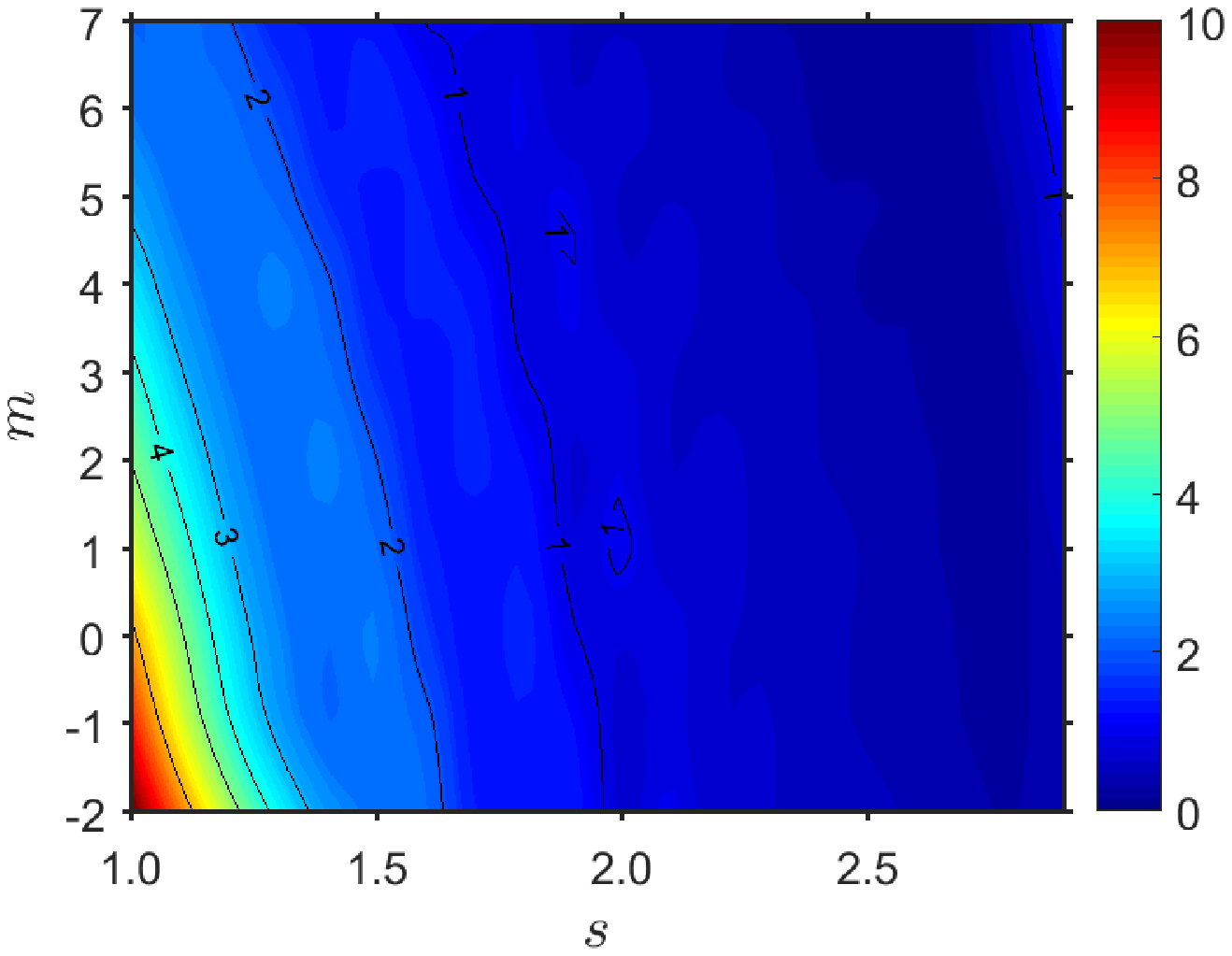} }}%
\subfloat[$\chi_{\rm min}^2$/d.o.f. as a function of $s$ and $E_\text{max}$, where $m$ is scanned.]{{\includegraphics[width=0.33\linewidth]{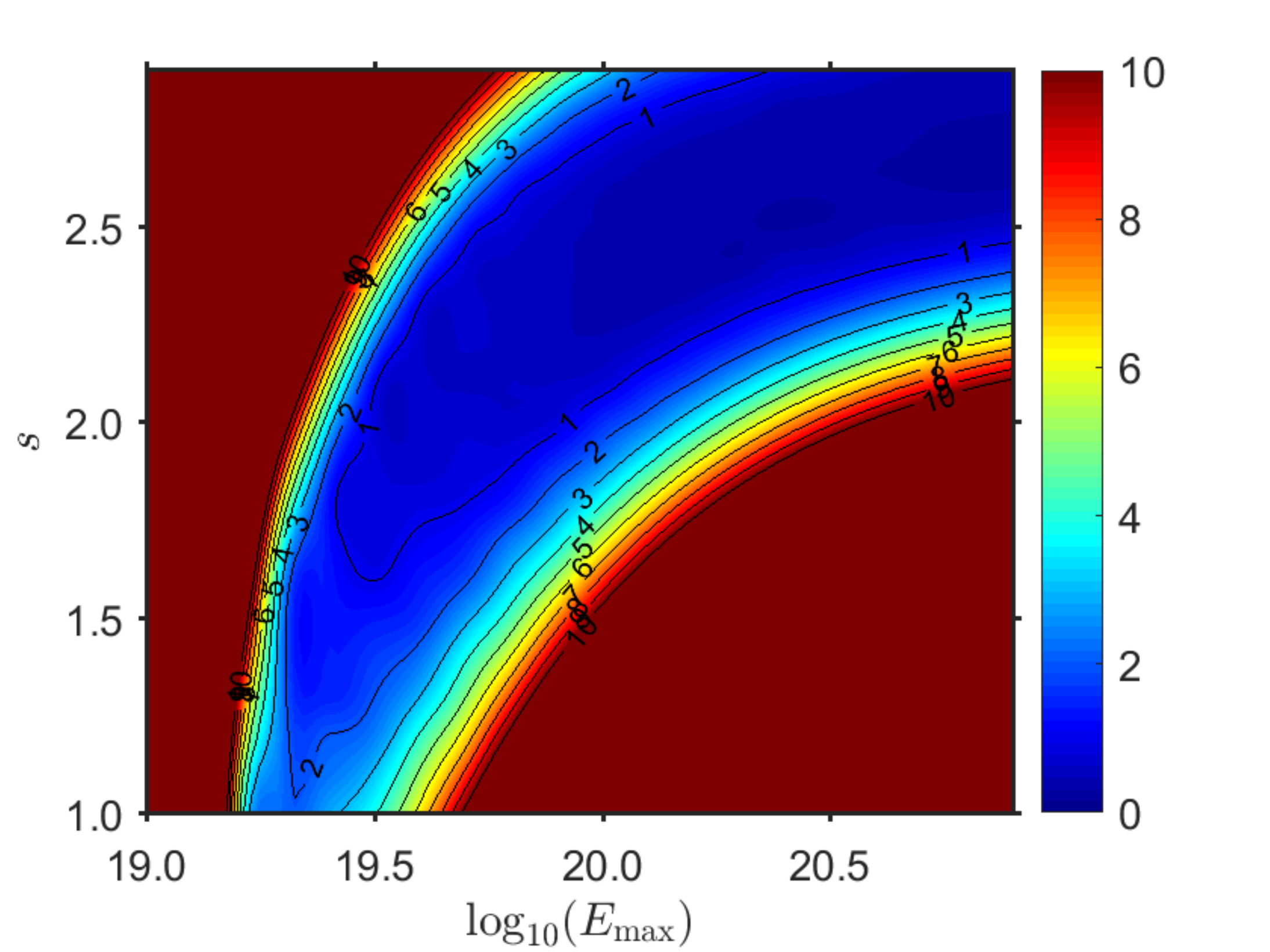} }}%
\caption{
Left plot shows our best-fit energy spectrum for the Auger data in the pure proton composition case. The differential simulated UHECR flux ($\Phi$) on Earth is multiplied by $E^3$. The best-fit parameters are $E_{p,\text{max}} = 10^{20.8}$~eV, $s = 2.7$, $m=5.0$, and $\delta_E =0.10$.
The middle plot shows the best-fit parameter space for the spectral index $s$ and the index of redshift evolution $m$. The solid contours indicate certain values of $\chi_{\rm min}^2$/d.o.f., where smaller ones give better fits to the spectrum. The right plot shows the best-fit parameter space for $s$ and $E_\text{max}$.}%
\label{fig:3DscanProton19}%
\end{figure*}

First, the results for the proton composition are shown in Fig.~\ref{fig:3DscanProton19}, as compared to Fig.~\ref{fig:3DscanProton}. 
With this energy fitting range, the best-fit parameters are $E_{p,\text{max}} = 10^{20.8}$~eV, $s = 2.7$, $m=5.0$, $\delta_E =0.10$ and $\chi^2/\rm d.o.f = 0.30$. Values of $\chi^2/\rm d.o.f$ are small, and the resulting spectrum is rather soft. 
However, the best-fit spectrum shown in Fig.~\ref{fig:3DscanProton19}(a) turns out to overshoot the experimental data at energies below the fitting energy range. This implies that some mechanism to suppress the lower-energy UHECR spectrum is necessary for this case to be physically viable. 
With such a narrower fitting range, we find a set of parameters that can better fit spectral data although the low-energy part of the best-fit spectrum tends to overshoot the observations. In Fig.~\ref{fig:3DscanProton19}, the areas in the parameter space that lead to good fits are extended, but the trend remains the same. The resulting energy generation rate densities also turn out to be comparable. 

\begin{figure*}
\centering
\subfloat[Spectral fit for He]{{\includegraphics[width=0.25\linewidth,trim={0 0cm 0 0},clip]{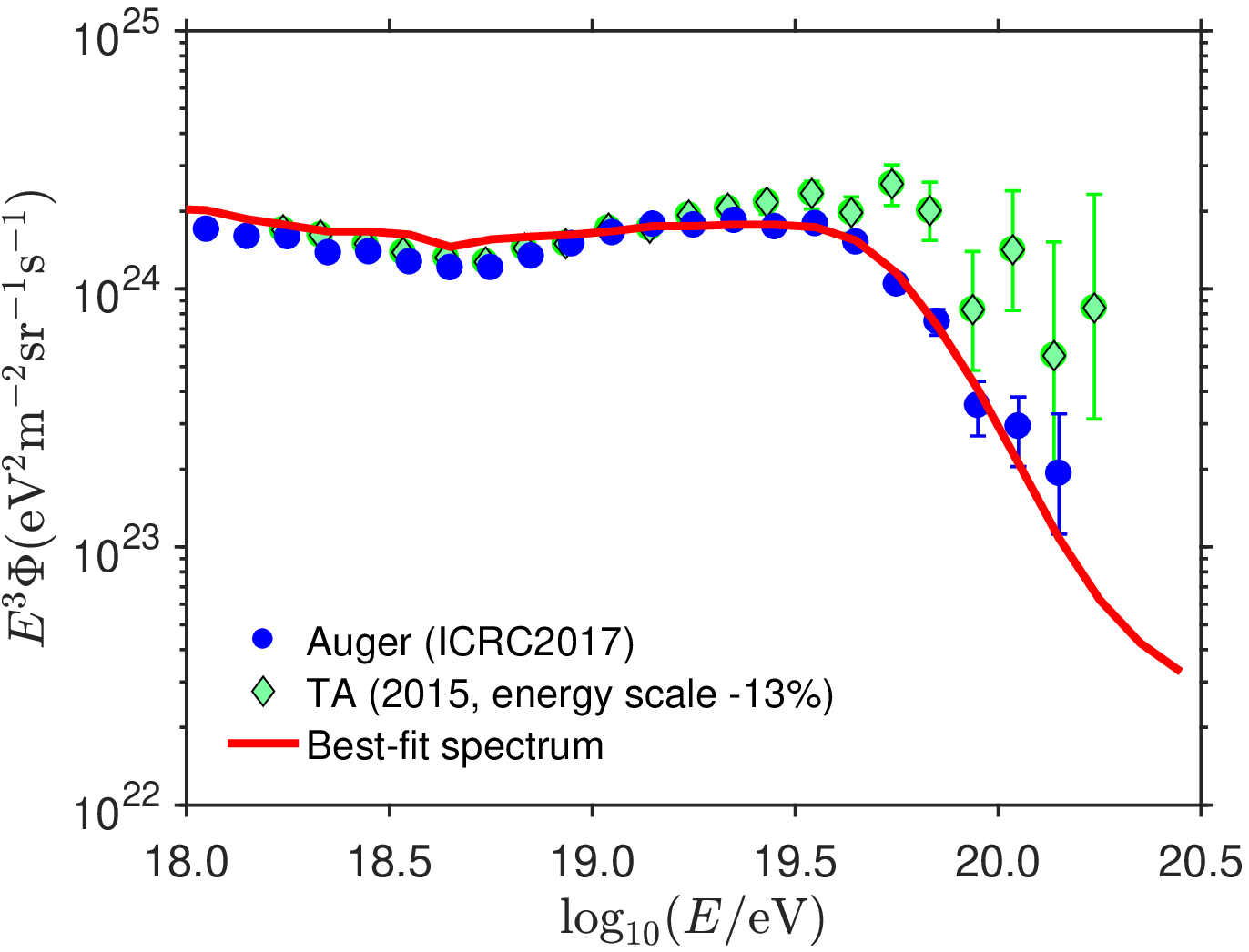} }}
\subfloat[Spectral fit for O]{{\includegraphics[width=0.25\linewidth]{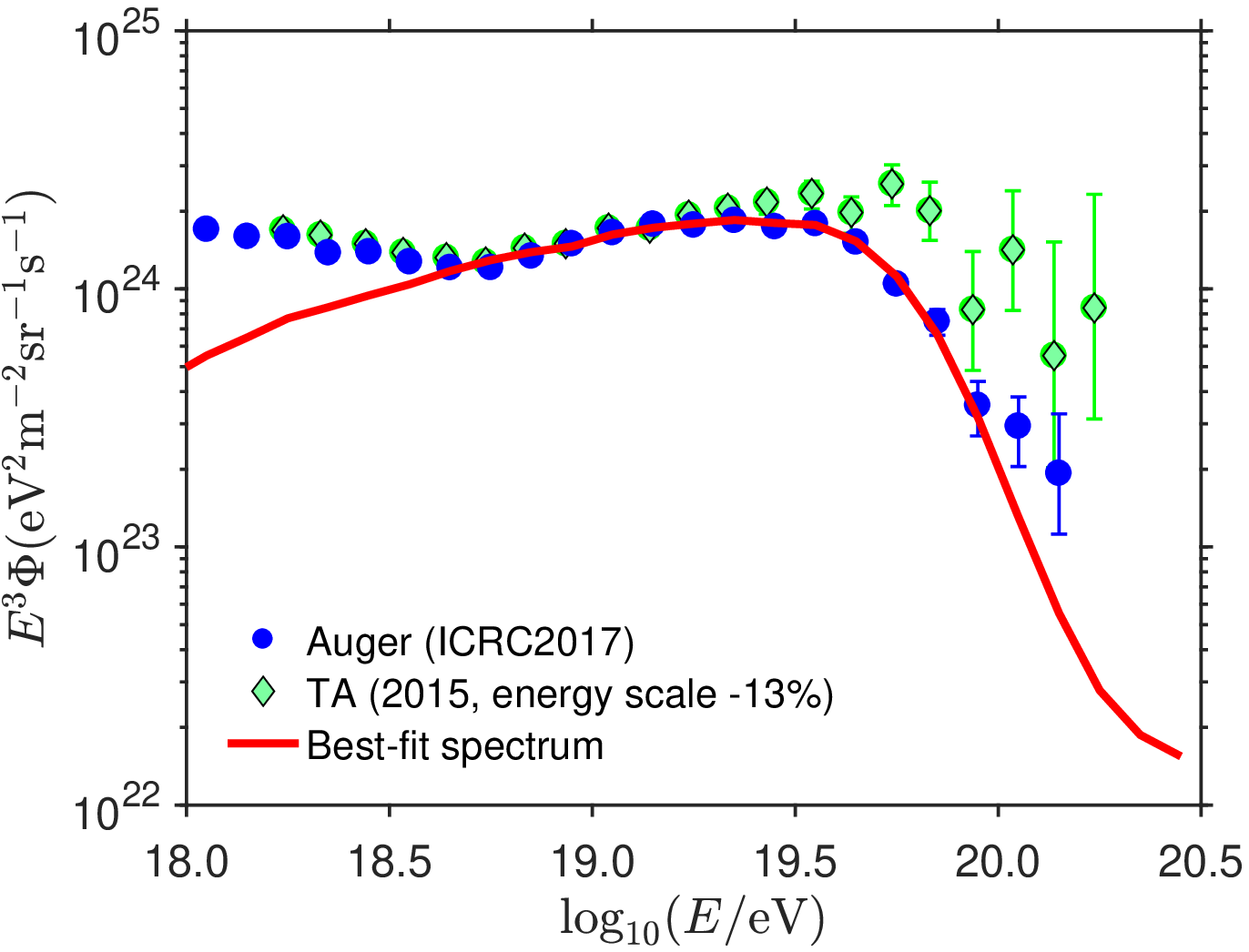} }}
\subfloat[Spectral fit for Si]{{\includegraphics[width=0.25\linewidth]{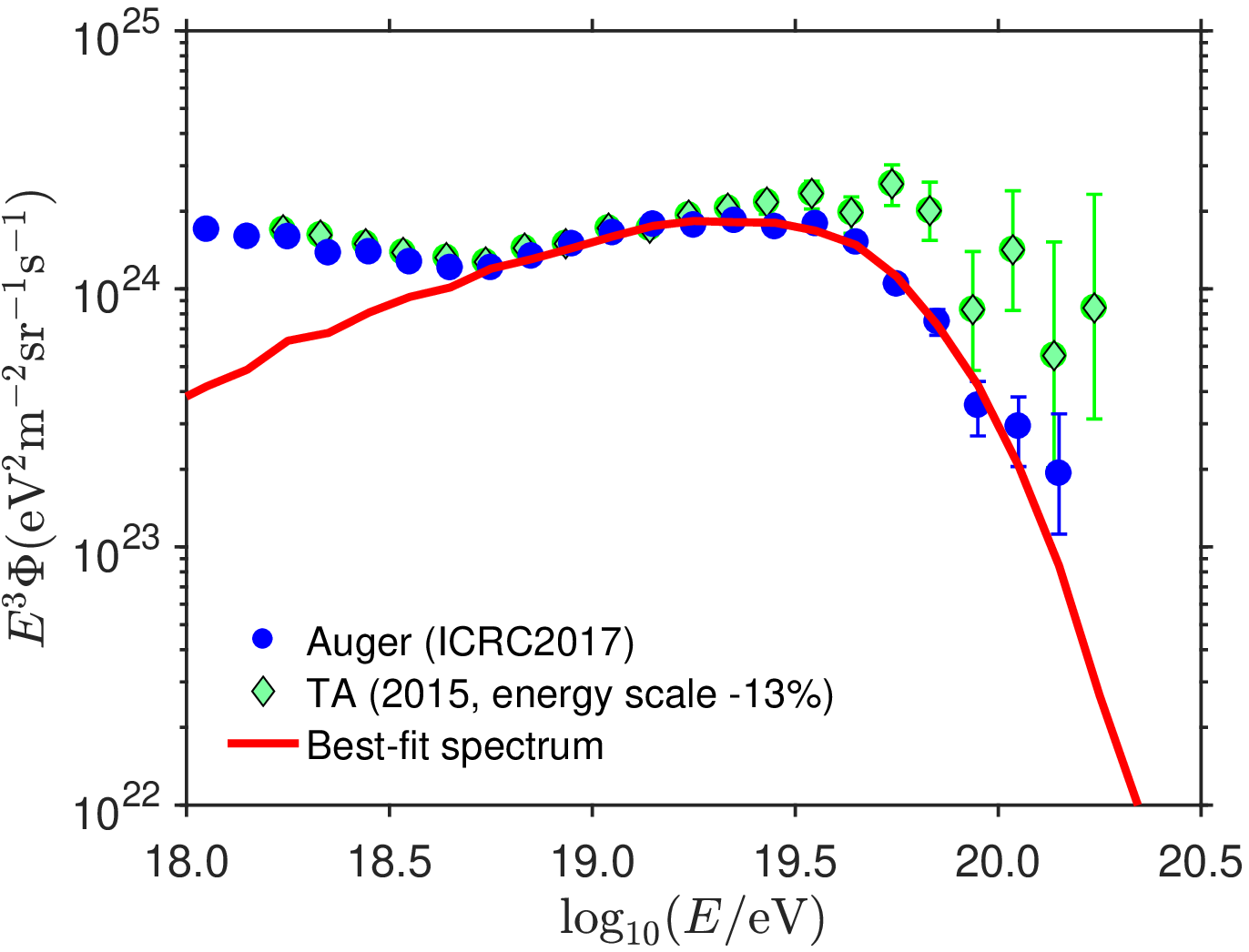} }}
\subfloat[Spectral fit for Fe]{{\includegraphics[width=0.25\linewidth]{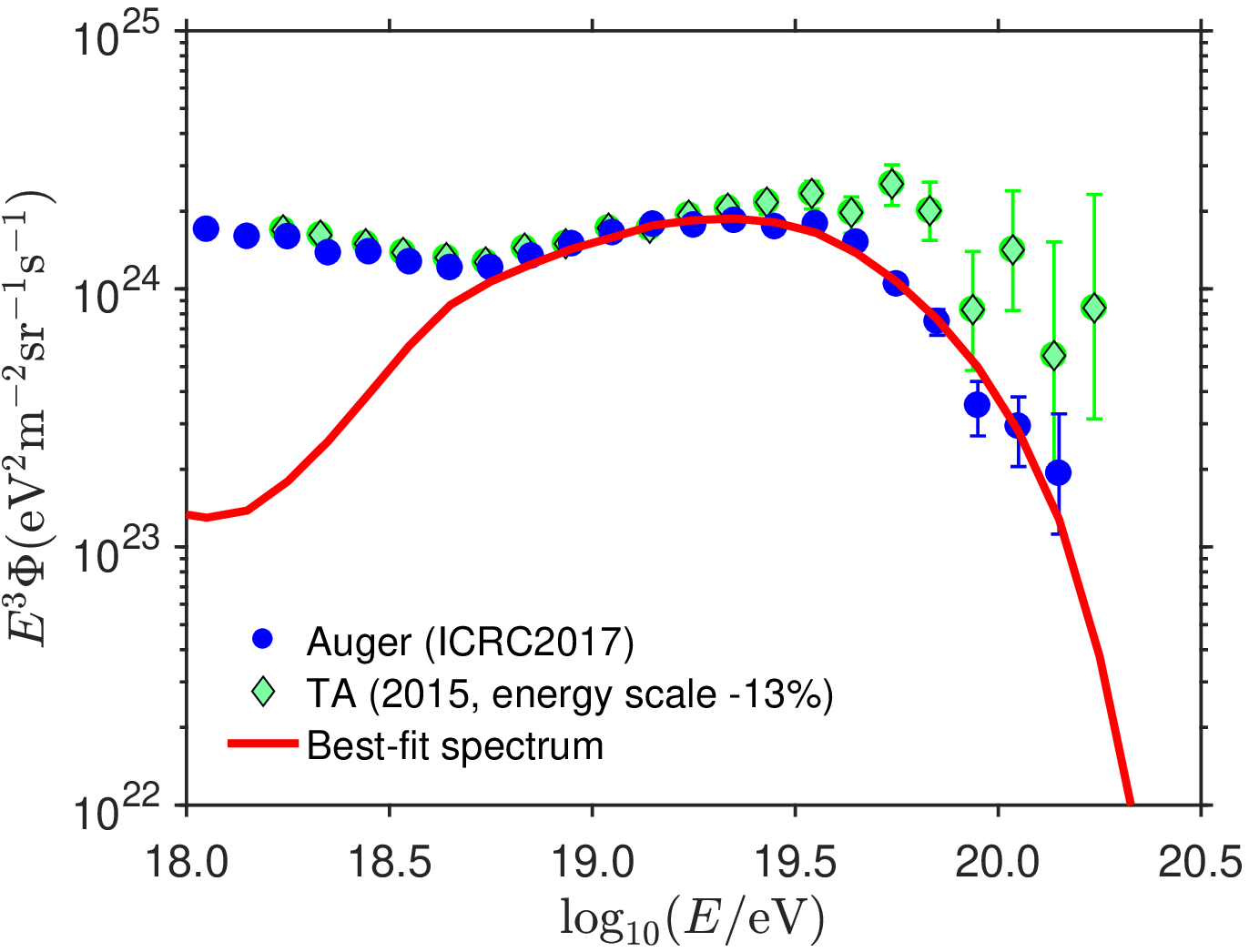} }}
\caption{Our best-fit spectra for pure He, O, Si and Fe injections, where the $X_{\rm max}$ data are not considered in the fits. Only $\chi_{\rm spec}^2$ is considered.}
\label{fig:NucleiSpecBestFit19}
\end{figure*}

\begin{figure*}%
\centering
\subfloat[He]{{\includegraphics[width=0.25\linewidth]{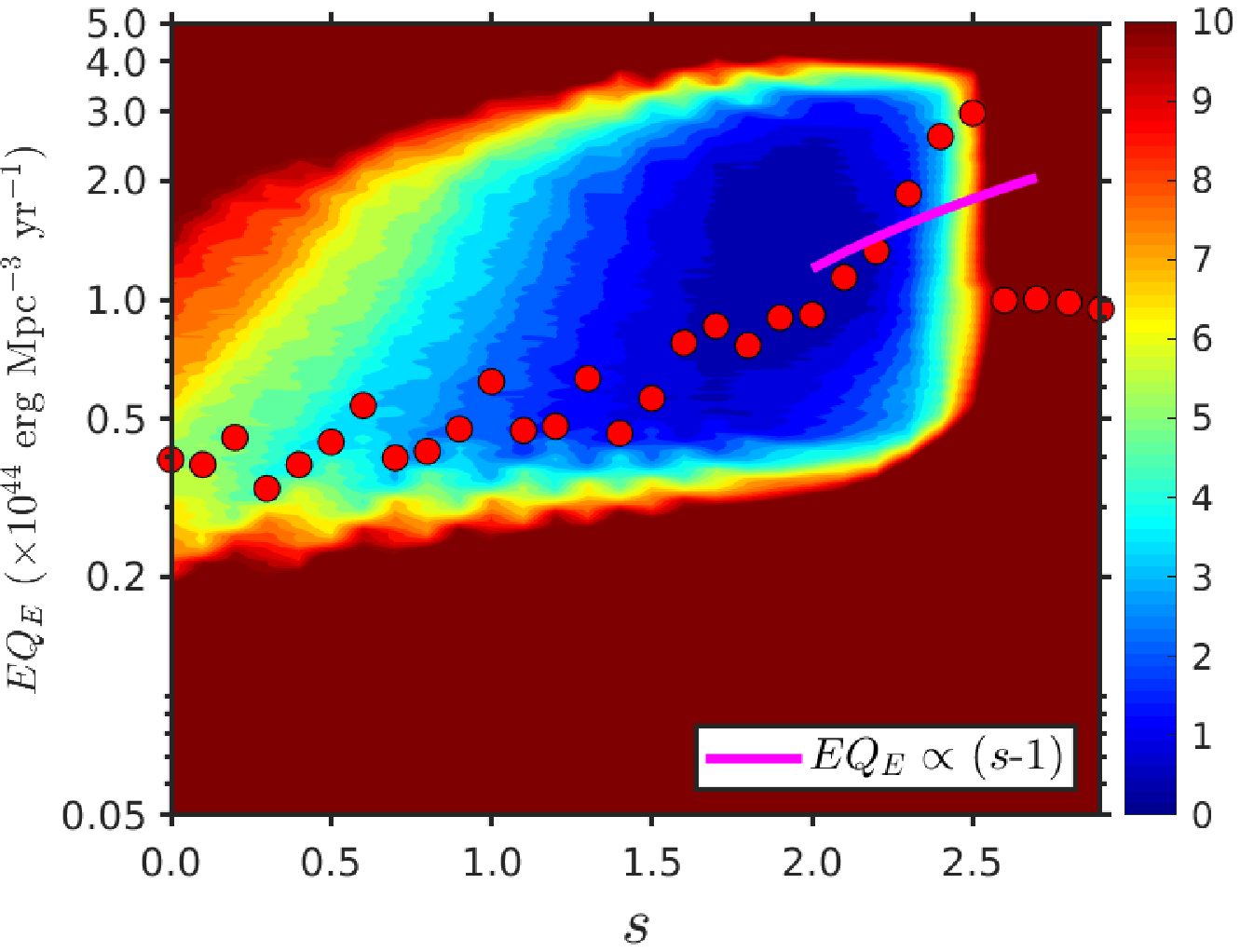}}}%
\subfloat[O]{{\includegraphics[width=0.25\linewidth]{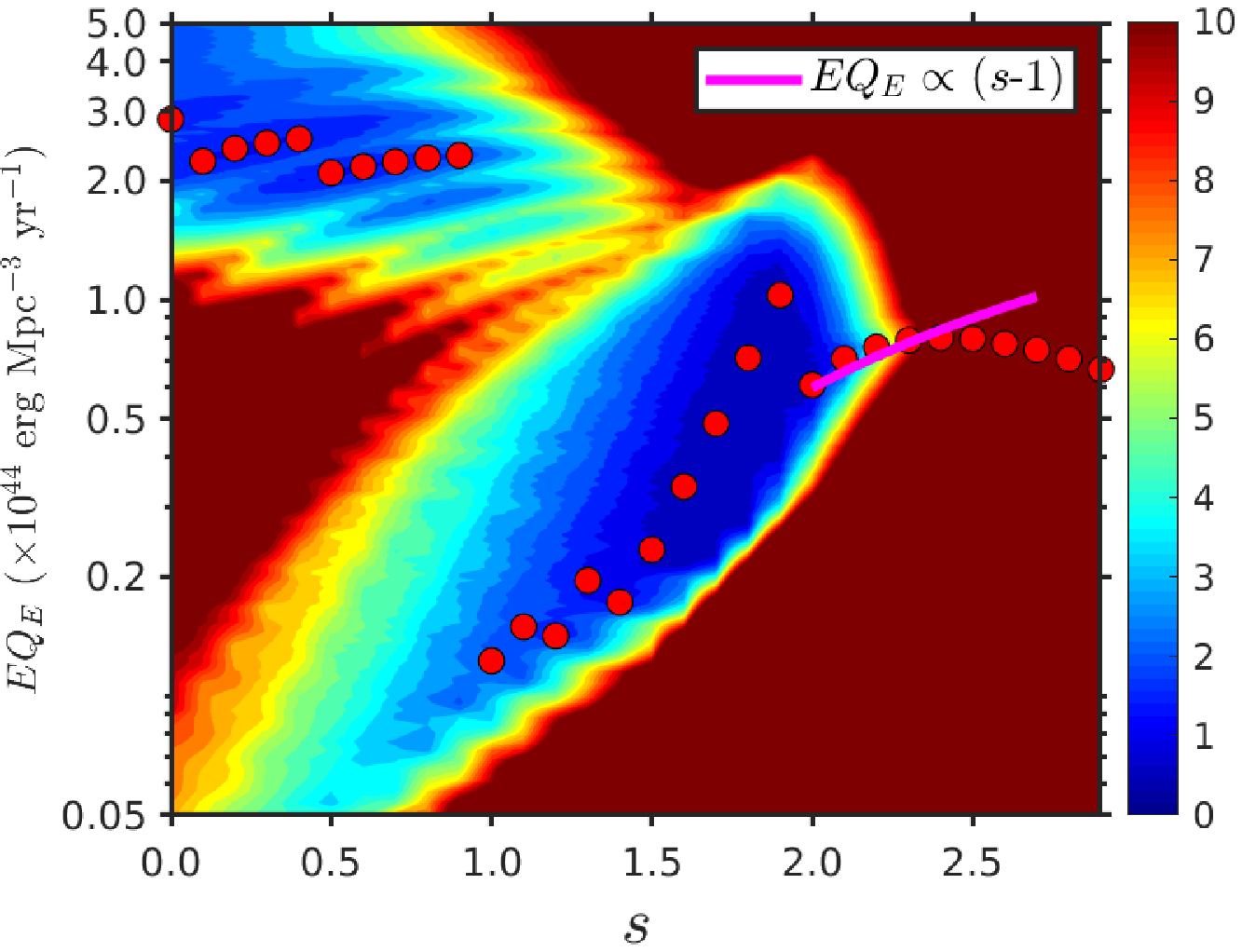}}}%
\subfloat[Si]{{\includegraphics[width=0.25\linewidth]{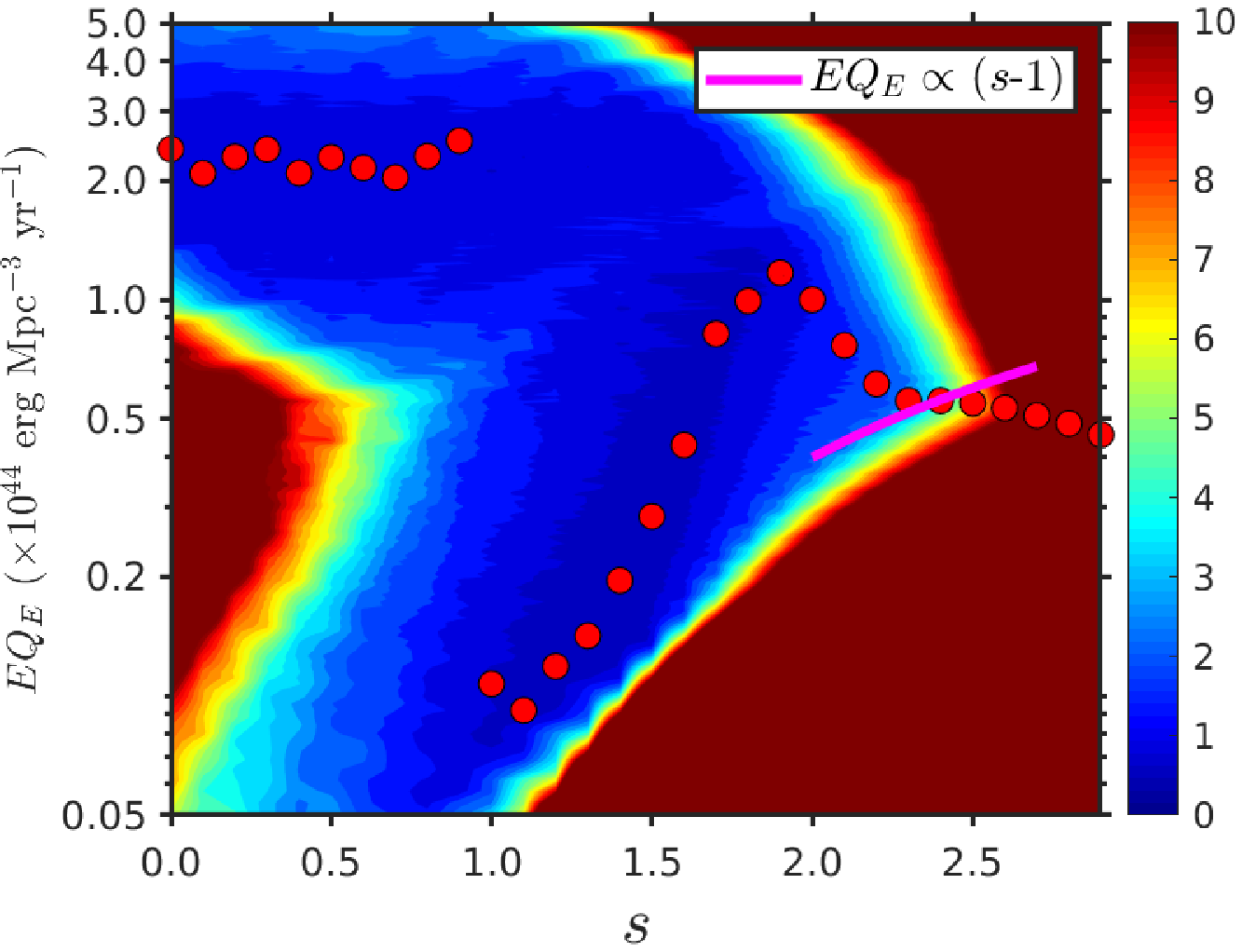}}}%
\subfloat[Fe]{{\includegraphics[width=0.25\linewidth]{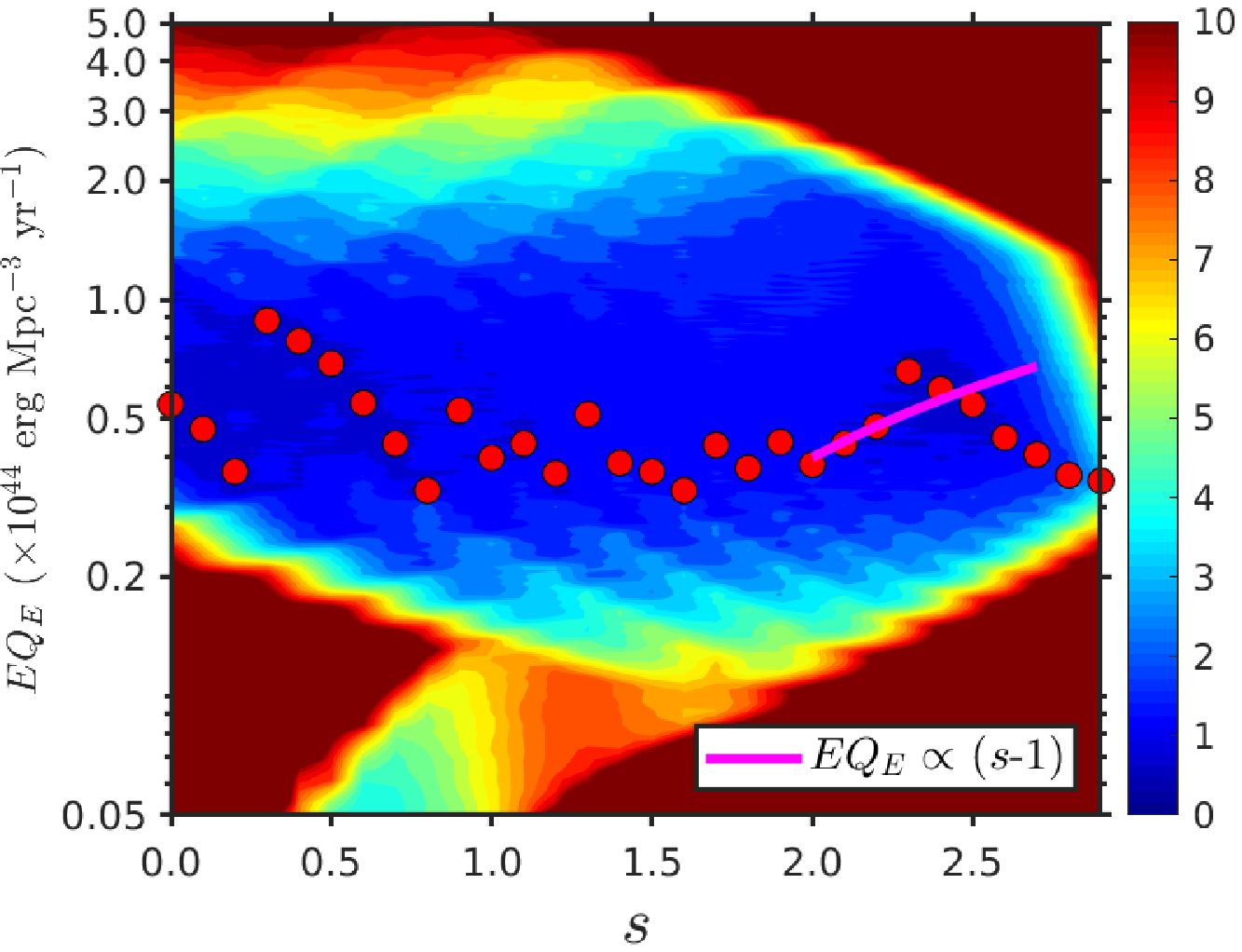} }}%
\caption{Best-fit energy generation rate density $EQ_E^{19.5}$ as a function of the spectral index $s$ (red solid dots), compared with the analytic dependence of $EQ_E^{19.5} \propto s -1 $ in the $s$ range of [2.0, 2.7]. As in Fig.~\ref{fig:injectionWithGammaProton}, the background contours show $\chi_{\rm min}^2$/d.o.f., and different species of nuclei at the sources are assumed in different subplots. The fitting energy range is from $10^{19.05}$~eV to $10^{20.15}$~eV.}%
\label{fig:injectionWithGammaNuclei19}%
\end{figure*}

The results for heavy nuclei are shown in Figs.~\ref{fig:NucleiSpecBestFit19} and \ref{fig:injectionWithGammaNuclei19}, as compared to Fig.~\ref{fig:NucleiSpecBestFit} and Fig.~\ref{fig:injectionWithGammaNuclei}, respectively. For heavier nuclei, the overshoot problem is basically absent as shown in Fig.~\ref{fig:NucleiSpecBestFit19}. When we do not consider the data points between $10^{18.45}$eV and $10^{19.05}$eV, the fit for iron injection is much improved as the value of $\chi^2/\rm d.o.f$ is reduced. But for the same reason of abandoning these data points, the fit in lower-energy range has larger discrepancies from the observations. 
As expected, the regions of low values of $\chi^2/\rm d.o.f$ in Fig.~\ref{fig:injectionWithGammaNuclei19} are expanded for each of the nuclear injection. Nevertheless, the relation of $EQ_E \propto (s-1)$ is not consistently valid for $s$ between 2.0 and 2.7. 
The energy generation rate densities for nuclei are similar to those in the other fitting range. Although the best-fit power-law index $s$ may change between the two fitting ranges, we can see the consistency in our results of $\chi_{\rm min}^2$ in parameter space scanning. 

\subsection{Spectral and composition fits}

If the $X_{\rm max}$ data are included in the fit for this energy range, values of $\chi^2_\text{total}$ for different combinations of $s$ and $m$ are similar. Although heavy nuclei like Fe do not have a good combined fit with the narrower fitting range, intermediate nuclei like oxygen and silicon have reasonably small values of $(\chi^2/\rm d.o.f)_\text{total}$. They not only give good combined fits but also favor smaller values of $m$, which is consistent with the previous work~\cite{Taylor:2015rla}. 

In summary, the choice of the fitting range impacts the goodness of fits: the fitting energy range from $10^{19.05}$eV to $10^{20.15}$eV is often accompanied by the overshoot problem for proton and light nuclei, but this issue could be resolved by the magnetic confinement of UHECRs in source environments or IGMFs. 
Also, this case gives not only more conservative estimates on the energy generation rate densities but also better combined fits for most of the nuclei when the $X_\text{max}$ data are taken into account. 
After all, our conclusions on the energy generation rate densities remain valid for different fitting ranges. 


\end{document}